%% file: CCGRID-MAIN.tex
\documentclass[conference]{IEEEtran}
\IEEEoverridecommandlockouts

\usepackage{graphicx}
\usepackage{subcaption}
\usepackage{booktabs}
\usepackage{multirow}
\usepackage{amsmath,amssymb,amsfonts}
\usepackage{algorithmic}
\usepackage{graphicx}
\usepackage{textcomp}
\usepackage{xcolor}
\def\BibTeX{{\rm B\kern-.05em{\sc i\kern-.025em b}\kern-.08em
    T\kern-.1667em\lower.7ex\hbox{E}\kern-.125emX}}
\begin{document}

\title{TopoSZp: Lightweight Topology-Aware Error-controlled Compression for Scientific Data

}

\author{\IEEEauthorblockN{Tripti Agarwal\IEEEauthorrefmark{1},
Sheng Di\IEEEauthorrefmark{2},
Xin Liang\IEEEauthorrefmark{3},
Zhaoyuan Su \IEEEauthorrefmark{4},
Yuxiao Li \IEEEauthorrefmark{5},
Ganesh Gopalakrishnan\IEEEauthorrefmark{1},\\
Hanqi Guo\IEEEauthorrefmark{5},
Franck Cappello\IEEEauthorrefmark{2}}
\IEEEauthorblockA{\IEEEauthorrefmark{1}University of Utah, Salt Lake City, UT, USA}
\IEEEauthorblockA{\IEEEauthorrefmark{2}Argonne National Laboratory, Lemont, IL, USA}
\IEEEauthorblockA{\IEEEauthorrefmark{3}University of Kentucky, Lexington, KY, USA}
\IEEEauthorblockA{\IEEEauthorrefmark{4}University of Virginia, Charlottesville, Virginia}
\IEEEauthorblockA{\IEEEauthorrefmark{5}The Ohio State University, Columbus, Ohio}

 tripti.agarwal@utah.edu, sdi1@anl.gov, xliang@uky.edu, acf7ea@virginia.edu, \\li.14025@buckeyemail.osu.edu, ganesh@cs.utah.edu, guo.2154@osu.edu, cappello@mcs.anl.gov

\thanks{Corresponding author: Sheng Di, Mathematics and Computer Science Division, Argonne National Laboratory, 9700 Cass Avenue, Lemont, IL 60439, USA}
}

\maketitle

\input{tex/0.abstract}

\input{tex/1.introduction}

\input{tex/2.related_work}

\input{tex/3.problem_formulation}

\input{tex/4.design}

\input{tex/5.performance_evaluation}

\input{tex/6.conclusion}

\bibliographystyle{IEEEtran}
\bibliography{references}

\end{document}

%% file: tex/0.abstract.tex
\begin{abstract}
Error-bounded lossy compression is essential for managing the massive data volumes produced by large-scale HPC simulations. While state-of-the-art compressors such as SZ and ZFP provide strong numerical error guarantees, they often fail to preserve topological structures (e.g., minima, maxima, and saddle points) that are critical for scientific analysis. Existing topology-aware compressors address this limitation but incur substantial computational overhead. We present TopoSZp, a lightweight, topology-aware, error-controlled lossy compressor that preserves critical points and their relationships while maintaining high compression and decompression performance. Built on the high-throughput SZp compressor, TopoSZp integrates efficient critical point detection, local ordering preservation, and targeted saddle-point refinement, all within a relaxed but strictly enforced error bound. Experimental results on real-world scientific datasets show that TopoSZp achieves 3×–100× fewer non-preserved critical points, introduces no false positives or incorrect critical point types, and delivers 100×–10,000× faster compression and 10×–500× faster decompression compared to existing topology-aware compressors, while maintaining competitive compression ratios.

\end{abstract}


%% file: tex/1.introduction.tex
\section{Introduction}
\label{sec:introduction}
The exponential growth of scientific data generation has created unprecedented challenges for data transmission, storage, and analysis. Modern simulations in climate modeling, combustion, cosmology, and materials science can produce terabytes of floating-point data per timestep—far exceeding what can be feasibly stored, transferred, or analyzed in full. As a result, data compression has become an essential component of modern scientific workflows. Among various techniques, error-bounded lossy compression has been widely developed and adopted because it can substantially reduce data size while providing explicit guarantees on reconstruction error. Traditional general-purpose lossy compressors (such as SZ~\cite{SZ1.4, SZ2, SZ3} and ZFP~\cite{ZFP}) focus primarily on pointwise error control—such as absolute or relative error bounds—or on statistical metrics such as peak signal-to-noise ratio (PSNR). However, these criteria do not directly address the preservation of topological structure, leaving a significant gap for scientific workflows in which topology is essential for downstream analyses such as feature tracking \cite{feature-tracking}, vortex and critical-point analysis \cite{helman1989visualizing}, and Morse–Smale complex extraction \cite{edelsbrunner2003morse}.

Topological structure is a critical feature that captures the fundamental geometric and organizational characteristics of scientific datasets. It describes how a scalar field is shaped through features such as minima, maxima, saddle points, and the connectivity patterns that relate them, forming the foundation of many higher-level scientific analyses. These structures enable researchers to identify coherent regions, detect transitions, quantify the evolution of physical phenomena, and extract domain-specific patterns of interest. For example, in climate science, topological descriptors support the tracking of vortices and atmospheric rivers~\cite{Muszynski2019AR}; in combustion, they delineate flame fronts and ignition kernels~\cite{Carr2003ContourTrees}; and in cosmology, they reveal density peaks, voids, and filamentary networks~\cite{sousbie2011disPerSE}. Such downstream analyses operate directly on the structure of the scalar field --- its arrangement of extrema~\cite{milnor1963morse}, the branching of contour trees~\cite{carr2003topological}, the organization of basins and ridges~\cite{gyulassy2008morse}, and the behavior and stability of critical points~\cite{edelsbrunner2002topological}. These properties are inherently topological: they depend not on precise numerical values, but on the subtle relationships, orderings, and connectivity that define the field’s geometric organization.

Preserving topological structure is essential for ensuring that compressed data remain scientifically reliable, yet guaranteeing such preservation is highly challenging --- even for error-bounded lossy compressors. This difficulty arises because topological structure depends on the relative ordering and connectivity of data values rather than their exact magnitudes. As a result, even small distortions introduced during compression can perturb these relationships, alter the underlying topology, and ultimately mislead downstream analyses. This insight motivated the primary design principle underlying our approach: preserving topological information requires a compression pipeline that is explicitly topology-aware.

Although prior efforts~\cite{TopoSZ,TopoA} have explored topology-preserving lossy compression methods, these approaches are often computationally expensive or achieve limited compression ratios, which hinders their routine use in large-scale scientific workflows. For scalar fields, topologically controlled schemes based on persistence diagrams and contour trees~\cite{TopoSZ,Soler2018TopologicallyCL} typically require global topological analysis and iterative reconstruction of sublevel sets to enforce constraints on critical points and their pairings. Recent topology-aware extensions of error-bounded compressors for scalar data, such as methods~\cite{TopoSZ} that augment SZ with contour-tree–guided constraints, further refine local upper and lower bounds and repeatedly adjust reconstructed values to eliminate topological inconsistencies. When the global error bound or persistence threshold is tight, these iterative correction passes can significantly reduce compression ratios and increase runtime. For vector and tensor fields, topology-preserving approaches~\cite{TFZ} that protect critical points or eigenstructure similarly derive per-sample error bounds from topological predicates (e.g., sign-of-determinant conditions) and invoke local repair or re-encoding until all constraints are satisfied. While effective at improving topological fidelity, these methods incur notable computational cost and often require complex, nontrivial implementations. Classic topology-preserving compression techniques for vector fields~\cite{Sign-of-determinant}, which are based on repeated edge collapses with local topology tests, likewise rely on fine-grained iterative operations and have primarily been demonstrated on relatively small two-dimensional domains. As a result, they are difficult to scale to modern scientific simulations producing multi-terabyte datasets.

In this paper, we propose a fast, topology-aware, error-controlled lossy compression method TopoSZp for scientific datasets, built on SZp \cite{hZCCL}, a state-of-the-art lightweight error-bounded compressor. Our core idea is to incorporate a suite of lightweight topology-preserving mechanisms directly into SZp’s quantization stage --- the primary source of data distortion in the compression pipeline. By regulating how values are quantized and selectively refining regions where topology is at risk, we aim to preserve essential structural features while retaining SZp’s efficiency and gracefully degraded compression ratios. We summarize the key contribution of TopoSZp as follows:
\begin{itemize}
    \item We develop a fast, topology-aware, error-controlled lossy compressor that preserves critical points and their relationships in large-scale scientific data. Our method introduces a relaxed but strictly enforced error bound accompanied by theoretical guarantees, enabling efficient compression while maintaining topological correctness.
    \item We carefully optimize the topological preservation pipeline by employing \textbf{lightweight critical point detection}, \textbf{applying extrema stencils to preserve maxima and minima}, and \textbf{performing RBF-based refinement for saddle points} using multithreaded OpenMP parallelism. This design significantly reduces the computational overhead typically associated with topology-aware compression.
    \item We perform a comprehensive evaluation on multiple real-world scientific datasets to demonstrate the effectiveness of our topology-aware compressor. Our results show that TopoSZp preserves topological critical points significantly better than traditional error-bounded lossy compressors, yielding \textbf{3×–100× fewer missing critical points}. In addition, TopoSZp achieves substantial performance gains, providing \textbf{100×–10,000× speedup in compression time} and \textbf{10×–500× speedup in decompression time} compared to other topology-aware compressors. Furthermore, TopoSZp also guarantees \textbf{zero false new critical points} and \textbf{zero false critical point types}, highlighting its strong topological fidelity. 
    
\end{itemize}

The remainder of this paper is organized as follows. Section~\ref{sec:related_work} reviews related work on error-bounded and topology-aware lossy compression. Section~\ref{sec:Problem} formulates the problem. Section~\ref{sec:design} presents the compression and decompression pipeline of TopoSZp. Section~\ref{sec:evaluate} reports experimental results and analysis on 5 HPC datasets. Finally, Section~\ref{sec:conclude} concludes the paper and discusses future work.

%% file: tex/2.related_work.tex
\section{Background and Related Work}
\label{sec:related_work}
This section begins with a review of topology-aware compression approaches, followed by a summary of existing error-bounded lossy compressors. We then provide a detailed discussion of the SZp compressor and its quantization procedure, which forms the foundation of our proposed method.

\subsection{Topology-aware compression}
The intersection of topology preservation and data compression has only recently gained attention. Although early efforts focused on mesh compression, far fewer methods address scalar field compression, where the core challenge is preserving topological structure without compromising compression ratio or error control.

Soler et al. \cite{Soler2018TopologicallyCL} introduced one of the first compressors to explicitly bound topological error. Building on this idea, TopoSZ \cite{TopoSZ} extends SZ-1.4 \cite{SZ1.4} to better preserve topological features. Gorski et al. \cite{TopoA} proposed a general wrapper (named as TopoA in our experiments) that enforces topological guarantees around existing compressors (SZ3, TTHRESH, ZFP). For vector and tensor data, Xia et al. \cite{Sign-of-determinant} used sign-of-determinant predicates to characterize critical points, and TFZ \cite{TFZ} preserves tensor-field–specific features while relying on SZ3/SPERR for numerical reduction. Additional efforts include preserving anatomical structures in medical images \cite{MEYERBASE2005383} and compressing chain complexes to retain multi-parameter persistence information \cite{Fugacci}.

In general, topology-aware compressors show promise, but still face challenges in speed, scalability, and integration with standard workflows --- motivating our design of TopoSZp.

\subsection{Error-bounded lossy compression}
Error-bounded lossy compression has become central to scientific data management, providing controllable reconstruction accuracy while significantly reducing storage and I/O overhead. Several compressors have been widely adopted across scientific domains. ZFP~\cite{ZFP} performs block-based bit-plane encoding, MGARD~\cite{MGARD} applies a hierarchical multigrid transform for progressive refinement, and TTHRESH~\cite{TTHRESH} uses tensor-train decomposition with coefficient thresholding for multidimensional fields. These methods deliver high compression efficiency with strict error guarantees, but primarily optimize throughput or compression ratio without explicitly preserving structural topological relationships within the data.

In parallel, the SZ family of compressors has evolved into one of the most widely used error-bounded frameworks for scientific workflows. The original SZ framework introduced prediction-based quantization with entropy coding. SZ3~\cite{SZ3} builds on SZ2~\cite{SZ2}, achieving higher compression ratios at comparable speeds using Lorenzo prediction~\cite{Lorenzo_prediction}, dynamic spline
interpolation, and entropy coding with lossless backends such as Huffman~\cite{Huffman} + GZIP~\cite{gzip1992}.
SZp (named as fZ-light in \cite{hZCCL}) increases throughput via kernel optimization and OpenMP parallelism;  
hoSZp \cite{hoSZp} and cuSZp \cite{cuSZp} extend it to heterogeneous systems and GPU. Other compressors, such as \cite{hZCCL} and \cite{SZOps}, balance speed and ratio, but SZp is currently the primary high-performance OpenMP-based version. Overall, despite their efficiency, existing error-bounded compressors do not preserve topological features, highlighting the need for our topology-aware TopoSZp design.

\subsection{SZp Compressor}
\label{sec:SZp}

In this subsection, we describe SZp in greater detail as essential background, since it serves as the foundational error-bounded lossy compression framework upon which our lightweight, topology-aware, error-controlled compression method is built. SZp is a lightweight, error-bounded lossy compressor designed for high-throughput floating-point data. Its simple quantization-based approach respects user-defined error bounds while minimizing computational overhead, making it well suited for large-scale scientific workflows that require fast compression and decompression.

The SZp compression pipeline consists of three stages:  \textbf{(1)~Quantization (Lossy Stage):} Each grid point value is quantized from floating point to an integer representation under a user-specified error bound, guaranteeing that the reconstructed value respects the given error constraint. 
\textbf{(2) Prediction:} A lightweight offset-based or neighbor-reuse strategy is used to predict the current data value. \textbf{(3) Fixed-Rate Byte Encoding:} The quantized integers are stored using a fixed number of bytes, avoiding heavy entropy coding stages such as Huffman, ANS, or arithmetic coding and enabling very fast encoding/decoding.

\textbf{Quantization.}  
Since quantization is the only lossy step in SZp, any loss of information during this stage can directly affect the reconstructed topology, potentially removing or altering critical points. In this section, we analyze this quantization step in detail and discuss its implications for preserving topological integrity.

\begin{figure}[!h]
    \centering
     \vspace{-0.8em}
    \includegraphics[width=\linewidth]{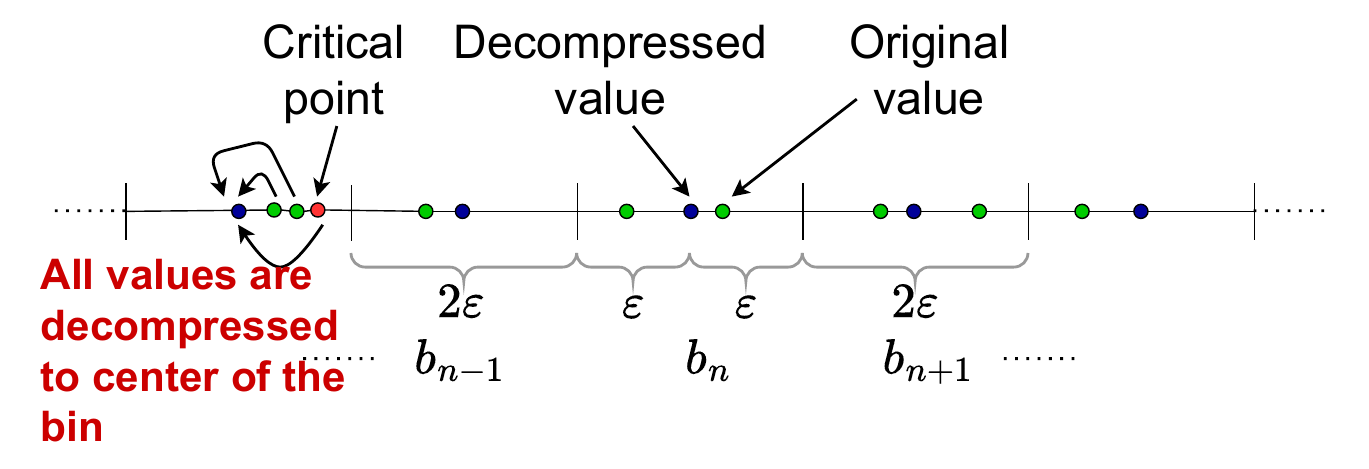}
    \vspace{-1.9em}
    \caption{Quantization encoder in SZp with error bound $\varepsilon$. Green dots represent the original samples, the red dot denotes a critical point, and the blue dot indicates the reconstructed value, corresponding to the center of the quantization bin.}
    \label{fig:quantization}
\end{figure}

Figure~\ref{fig:quantization} illustrates the SZp quantization encoder under a user-defined error bound $\varepsilon$, where $b_i,\ \forall i \in [0,m)$; $m \in \mathbb{N}$, denote the bin indices.

SZp follows a simple pipeline, in which the quantization step converts the prediction residual into a discrete bin index (integer value). This index is calculated as: $q_a = \lfloor \frac{a + \varepsilon}{2.\varepsilon}\rfloor$, where $a$ denotes the data value being quantized (i.e., the floating-point value at a grid point) and $q_a$ is its quantized bin index. During decompression, the inverse operation reconstructs the value by mapping the bin index back to the center of its quantization interval: $\hat{a} = q_a.(2\varepsilon) - \varepsilon$, ensuring that the reconstruction error remains bounded by $\varepsilon$. Note that the reconstructed value $\hat{a}$ is not equal to the original value $a$, but is guaranteed to satisfy $|\hat{a} - a| \leq \varepsilon$. This uniform linear quantization enables efficient entropy coding and forms the core mechanism used in SZp for error-controlled compression.

However, because multiple residual values within a $2\varepsilon$ range are mapped to the same bin center, small but meaningful variations in the original data may be flattened/removed during quantization (also shown in Figure~\ref{fig:quantization}). Consequently, critical points and their relative order can change or disappear, leading to a potential loss of topological structure. We further discuss how critical points are lost in detail in Section~\ref{sec:Problem}.

%% file: tex/3.problem_formulation.tex
\section{Problem formulation}
\label{sec:Problem}
We formulate the research problem as follows. Let the raw data be a scalar field $\mathcal{D}: \Omega \rightarrow \mathbb{R}$, defined on a 2D structured grid $\Omega = \{0,\ldots,n_x-1\} \times \{0,\ldots,n_y-1\}$. For a critical point preserving error-bounded lossy compressor (i.e., TopoSZp), it would perform compression of the data $\mathcal{D}$ leading to a compressed data stream (denoted by $c_t$), whose corresponding decompressed data is denoted by $\mathcal{\hat{D}}_t$, which satisfies the bound $|\mathcal{D}(x,y) - \mathcal{\hat{D}}_t(x,y)| \le \varepsilon, \forall (x,y) \in \Omega$, while aiming to preserve the critical topological points of the original field.

\subsection{Failure from Local Flattening (Critical-Point loss)}
\label{sec:problem_A}
\begin{figure}[!h]
    \centering
     \vspace{-0.8em}
    \includegraphics[width=\linewidth]{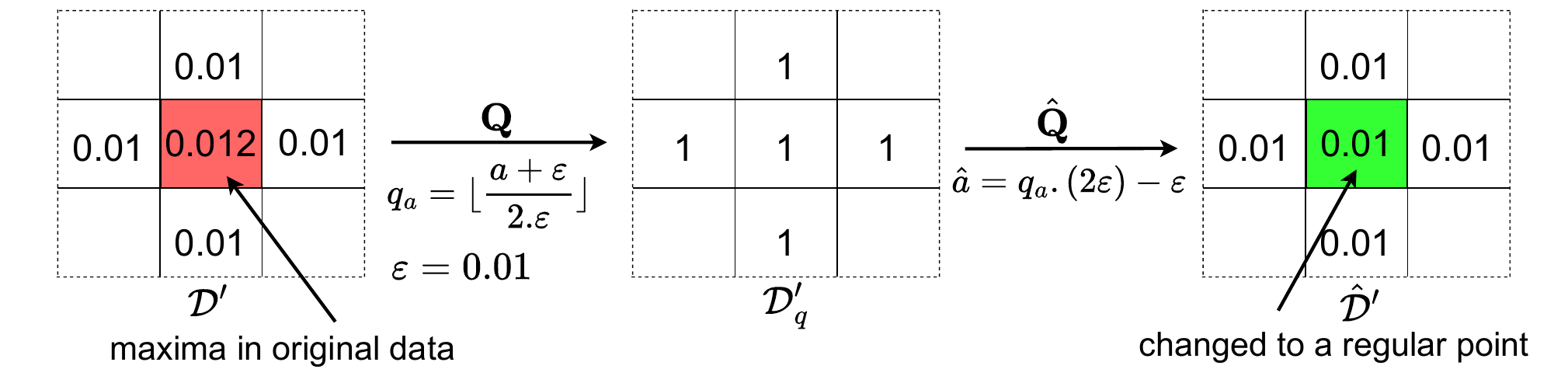}
    \vspace{-1.5em}
    \caption{Quantization/dequantization removes the maxima  despite pointwise error guarantees.}
    \label{fig:problem1}
\end{figure}

A general error-bounded lossy compressor (such as SZp) guarantees pointwise accuracy but not 
\textbf{topological fidelity}. Even when every value satisfies the numerical constraint, small perturbations introduced during quantization may significantly \textbf{alter the local topological structure}. Figure~\ref{fig:problem1} illustrates this effect for a small subsection (denoted $\mathcal{D'}$) of the entire scalar field $\mathcal{D}$. In $\mathcal{D'}$, the center value $0.012$ forms a clear maximum ($\mathbf{M}$) with all four neighbors having lower values as $0.01$. For an error bound of $\varepsilon = 0.01$, this peak can be flattened or reversed, leading to a loss or change of critical points.

This is because, during quantization ($\mathbf{Q}$), the center value and its neighbors are assigned to a discrete bin index using $q_a = \lfloor \frac{a + \varepsilon}{2.\varepsilon}\rfloor$ (as mentioned previously in~\ref{sec:SZp}). In the subsequent dequantization step ($\hat{\mathbf{Q}}$), the reconstructed value $\hat{a}$ is replaced by the representative value of that bin (computed via the inverse quantization mapping). In our example in Fig~\ref{fig:problem1}, all values are mapped to $q_a =1$ and $\hat{a} = 0.01$. Although $\hat{a}$ remains within the error bound $\varepsilon = 0.01$, this mapping may destroy the strict ordering required to classify the center as a maximum $\mathbf{M}$. As a result, the original $\mathbf{M}$ is lost and misclassified as a regular point. Similar ordering inversions can also disrupt the identification of other critical structures—including minima ($\mathbf{m}$), saddles ($\mathbf{s}$), and maxima ($\mathbf{M}$)—highlighting that local topological features may not be preserved even under strict pointwise error guarantees.

\subsection{Classification of Topological Errors}
\label{sec:problem_B}
In principle, topological inconsistencies fall into three categories: \textbf{false positives (FP), false negatives (FN), and false types (FT)} \cite{TopoSZ}. However, SZp enforces a strict pointwise error bound and its quantization/dequantization process is monotone (i.e., $a_1 < a_2 \implies \hat{a}_1 \leq \hat{a}_2$). This implies that reconstruction cannot introduce new critical points that were not present in the original field (eliminating FP). This is because if $p$ is not a maximum in $\mathcal{D}$, then $\exists q\in N(p)$ where $N$ represents the neighbor of $p$ with $\mathcal{D}(q)\ge\mathcal{D}(p)$. Monotonicity implies 
$\hat{\mathcal{D}}(q)\ge\hat{\mathcal{D}}(p)$, so $p$ cannot become a new maxima. Similarly, a new minima or saddle can never occur in SZp decompressed data. Hence, \textbf{FP is impossible.} 
Similarly, SZp cannot alter the ordering strongly enough to transform a true maxima into a minima or saddle, or vice versa. This is because if $p$ is a strict maxima, then $\mathcal{D}(p) > \mathcal{D}(q)$ for all $q \in N(p)$. After reconstruction $\hat{\mathcal{D}}(p) \geq \hat{\mathcal{D}}(q)$, therefore $p$ cannot become minima or saddle. Similarly, minima or saddle can also never change their type. Hence, \textbf{FT is impossible}.
The only remaining form of topological distortion is therefore false negatives (FN), in which a true critical point in the original field is misclassified as a regular point after quantization ($\mathbf{Q}$) and dequantization ($\hat{\mathbf{Q}}$). This is possible when the difference between $p$ and a neighbor satisfies $\mathcal{D}(p) - \mathcal{D}(q) < 2\varepsilon$. Then, quantization can map them into the same bin, leading to an FN case (example shown in Fig.~\ref{fig:problem1}). These \textbf{FN cases} serve as the dominant failure mode in topology preservation and directly motivate the corrective strategies employed by TopoSZp.

\subsection{Failure of Relative-order Collapse}
\label{sec:problem_C}
\begin{figure}[!h]
    \centering
     \vspace{-1.2em}
    \includegraphics[width=\linewidth]{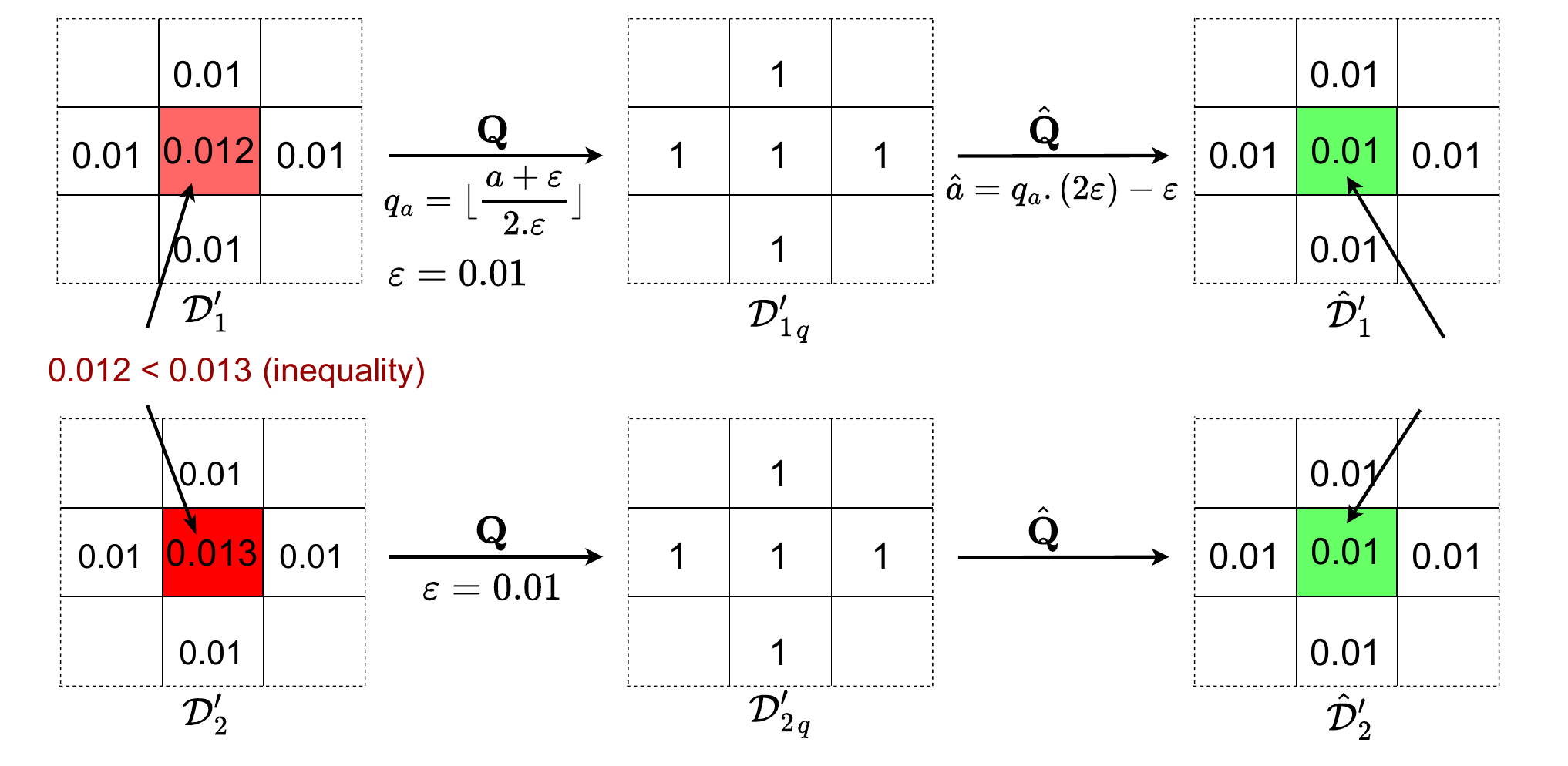}
    \vspace{-1.5em}
    \caption{Relative-order loss after quantization and subsequent dequantization.}
    \label{fig:problem2}
    \vspace{-1em}
\end{figure}

The next challenge arises from the loss of \textbf{relative ordering} among critical points. Even when all values satisfy the pointwise error bound, quantization ($\mathbf{Q}$) and subsequent dequantization ($\hat{\mathbf{Q}}$) may still alter the \textbf{local value relationships} that determine the topological structure. Figure~\ref{fig:problem2} demonstrates this effect using two local subsections of the scalar field, denoted $\mathcal{D}'_1$ and $\mathcal{D}'_2$. In $\mathcal{D}'_1$, the center value $0.012$ forms a local maxima ($\mathbf{M_1}$) relative to its neighbors, and in $\mathcal{D}'_2$, the center value $0.013$ similarly forms a maxima ($\mathbf{M_2}$). Although both $\mathbf{M_1}$ and $\mathbf{M_2}$ are valid critical points in the original field $\mathcal{D}$, their distinct magnitudes imply that $\mathbf{M_2}$ is globally more significant than $\mathbf{M_1}$, i.e. $\mathbf{M_2} > \mathbf{M_1}$.

However, after quantization ($\mathbf{Q}$) and dequantization ($\hat{\mathbf{Q}}$) with an error bound of $\varepsilon = 0.01$, their reconstructed values may collapse to the same bin representatives (e.g., both becoming 0.01 after $\hat{\mathbf{Q}}$). This not only risks turning the maxima into regular points (as discussed previously in section~\ref{sec:problem_A}) but also eliminates the relative inequality between them. Thus, even if the Critical-point loss  problem (\ref{sec:problem_A}) is addressed and both $\mathbf{\hat{M}_1}$ and $\mathbf{\hat{M}_2}$ remain classified as maxima, the reconstruction would satisfy $\mathbf{\hat{M}_1} = \mathbf{\hat{M}_2}$, erasing the original ordering $\mathbf{M}_1 < \mathbf{M}_2$. This loss of order can affect global topological reasoning, hierarchy construction, and assessment of important features necessary for topological preservation.

Together, these issues reveal a key limitation of relying solely on pointwise error bounds for topology preservation. Even when all values remain within $\varepsilon$, quantization can remove true critical points (false negatives (FN)), weaken saddle structures, and collapse the relative ordering between extrema. 
Since these relationships govern gradient flow and structural connectivity, their disruption leads to topological inconsistency in the reconstructed 
field. Preserving topology, therefore, requires maintaining both critical point classification and local ordering during compression, with selective refinement in sensitive regions. This motivates \textbf{TopoSZp}, which augments SZp with 
lightweight metadata and correction strategies to preserve numerical accuracy and topological structure.

Section~\ref{sec:design} presents the design strategies adopted in TopoSZp to address the challenges discussed.

%% file: tex/4.design.tex
\section{Design Overview}
\label{sec:design}

This section introduces the design of TopoSZp, a topology-aware extension of SZp that preserves critical points and their relationships under a relaxed yet strictly error-bounded framework, while maintaining high compression efficiency and fast compression/decompression performance.

\subsection{Compression Pipeline}

TopoSZp is an error-bounded floating-point compressor consisting of four sequential stages: Critical Point Detection and Relative Positioning (\textbf{CD} + \textbf{RP}),  Quantization (\textbf{QZ}), Blocking and Decorrelation (\textbf{B} + \textbf{LZ}), and Fixed-Length Byte Encoding (\textbf{BE}). The latter three stages (i.e., \textbf{QZ}, \textbf{B} + \textbf{LZ},  \textbf{BE}) follow the standard SZp workflow (Section~\ref{sec:SZp}), ensuring competitive compression ratio and throughput. The novelty of the TopoSZp compression pipeline lies in the first stage, \textbf{CD} + \textbf{RP}, where we explicitly extract topological information before lossy quantization. This step ensures that critical features are preserved even if surrounding values undergo significant rounding within the error bound.

    \textbf{(1) Critical Point Detection (CD).} Given the scalar field $\mathcal{D}: \Omega \rightarrow \mathbb{R}$, we classify each point $p \in \Omega$ by comparing it to its 4 neighbors top (t), bottom (d), left (l), and right (r). Note that the corner points use two neighbors, and the edge points use three. A point is categorized according to the following criteria:
    \begin{itemize}
        \item \textit{Minima} (\(\mathbf{m}\)): all available neighbors \((t,d,l,r)\) have strictly higher values than \(p\);  
        \[ 
            \mathcal{D}(p) < \mathcal{D}(n), \forall n \in \{t,d,l,r\}
        \] 
        \item \textit{Maxima} (\(\mathbf{M}\)): all available neighbors \((t,d,l,r)\) have strictly lower values than \(p\); 
        \[
            \mathcal{D}(p) > \mathcal{D}(n), \forall n \in \{t,d,l,r\}
        \]
        \item \textit{Saddle} (\(\mathbf{s}\)): two opposite neighbors (e.g., \(t,d\)) are higher and the other two (e.g., \(l,r\)) are lower than \(p\), or vice versa;
        \[
            \mathcal{D}(p) < \mathcal{D}(n), \forall n \in \{t,d\}, \mathcal{D}(p) > \mathcal{D}(n), \forall n \in \{l,r\}
        \]
        \item \textit{Regular} (\(\mathbf{r}\)): otherwise.
    \end{itemize}
    
\begin{figure}[!h]
        \centering
        \vspace{-0.8em}
        \includegraphics[width=\linewidth]{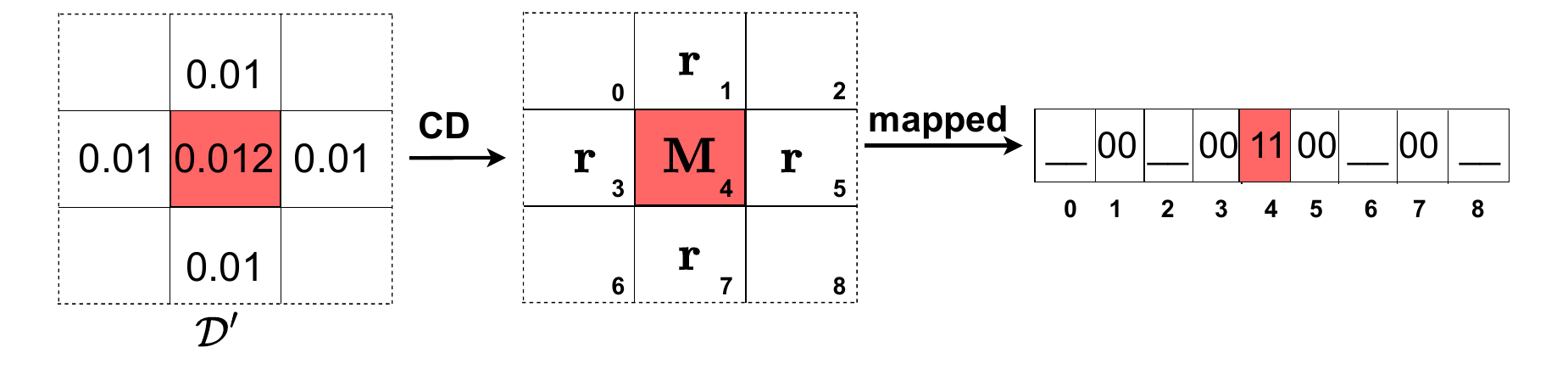}
        \vspace{-1.5em}
        \caption{Binary mask to represent the 2-bit stream to represent the type of point in the data. \_\_ also represents the 2-bit binary representations of the point type (i.e., 00, 01, 10, or 11) stored at their respective locations.}
        \label{fig:CD}
    \end{figure}

    Applying this classification to all grid points yields a label map with four categories: \(\{\mathbf{m}\), \(\mathbf{s}\), \(\mathbf{M}\),  \(\mathbf{r}\}\). Each class is encoded compactly using two bits: $\mathbf{r} = 0 \rightarrow 00$, $\mathbf{m} = 1 \rightarrow 01$, $\mathbf{s} = 2 \rightarrow 10$, $\mathbf{M} = 3 \rightarrow 11$. We show a small example of how the 2-bit binary mask looks like in Figure~\ref{fig:CD}. This binary mask forms the foundation for critical point tracking and is later used to recover lost extrema and saddles during decompression.
    
    \textbf{(2) Relative Positions (RP).}
    Although the CD stage ensures the identification of critical points, quantization may still distort the \emph{relative ordering} of the scalar values. This means that if the critical points remain correctly classified as minima~($\mathbf{m}$), saddles ($\mathbf{s}$), or Maxima ($\mathbf{M}$), the local perturbation introduced by quantization can flatten or reorder values, that can eventually break the original comparitive relationship (which we explained previously in  Section~\ref{sec:problem_C}). Such inversions directly affect gradient flow and connectivity of topological structures.

\begin{figure}[!h]
        \centering
         \vspace{-1.2em}
        \includegraphics[width=\linewidth]{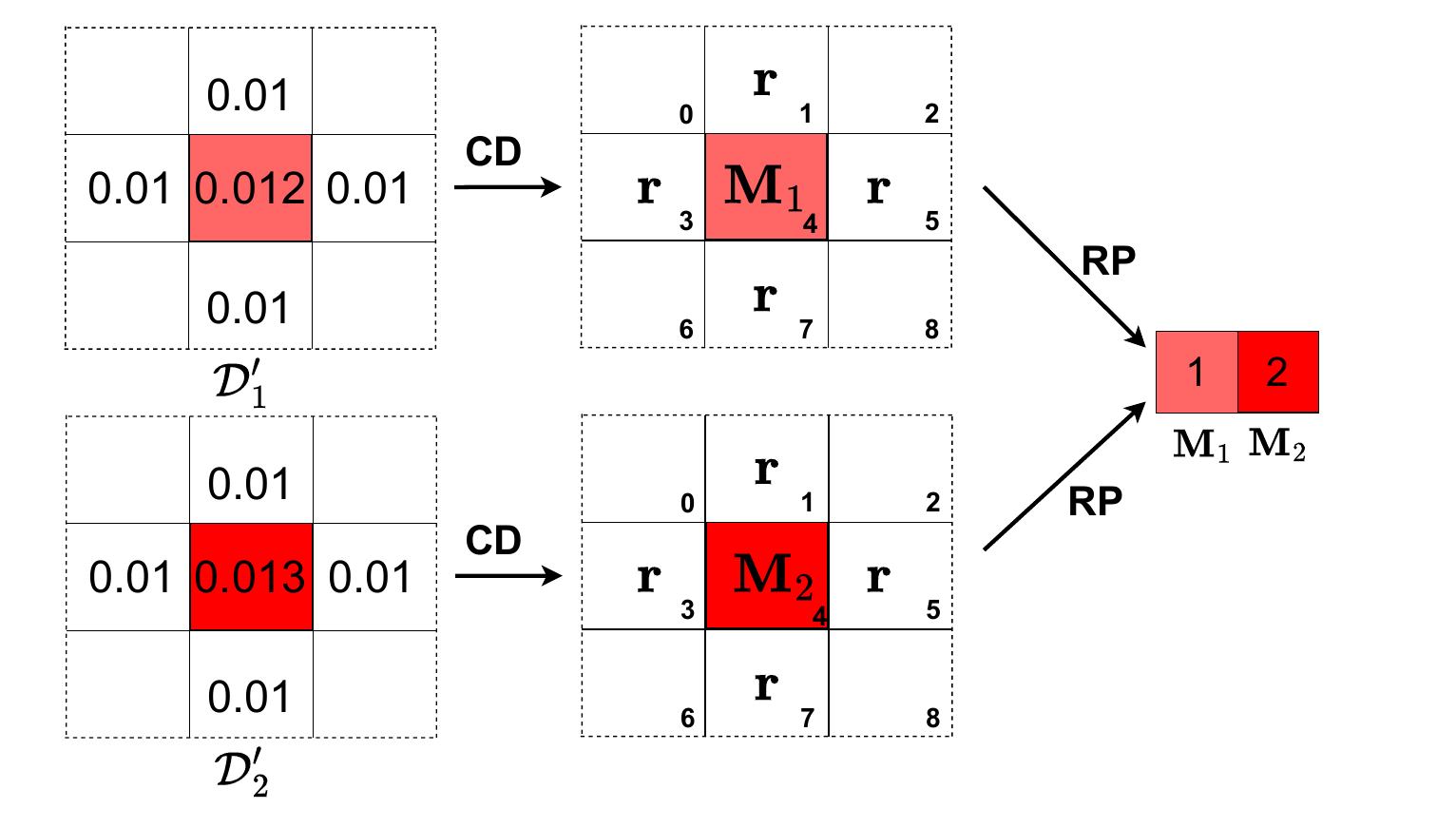}
        \vspace{-1.9em}
        \caption{Relative ordering stored for critical point $\mathbf{M_1}$ and $\mathbf{M_2}$ falling in same qunatization bin $1$. Since $\mathbf{M_1} < \mathbf{M_2}$ therefore its location is marked as 1 and $\mathbf{M_2}$ as 2.}
        \label{fig:RF}
    \end{figure}

    To address this, the RP stage captures minimal yet sufficient information to preserve relative ranking among critical points after compression. For each detected critical point $p$ falling in a particular quantization bin, we store small metadata indicating its value ranking (in the form of integer values) among other critical points falling in the same bin. We show an example of how the relative ordering in maintained in Figure~\ref{fig:RF}. During decompression, this stored ordering information is used to selectively adjust the reconstructed values to restore consistent relationships between critical points, preventing collapse events such as those shown in Section~\ref{sec:problem_C}. As a result, TopoSZp maintains both critical point classification and their topological hierarchy, enabling faithful reconstruction of structural relationships in the scalar field.

\begin{figure}[!h]
        \centering
         \vspace{-0.8em}
        \includegraphics[width=\linewidth]{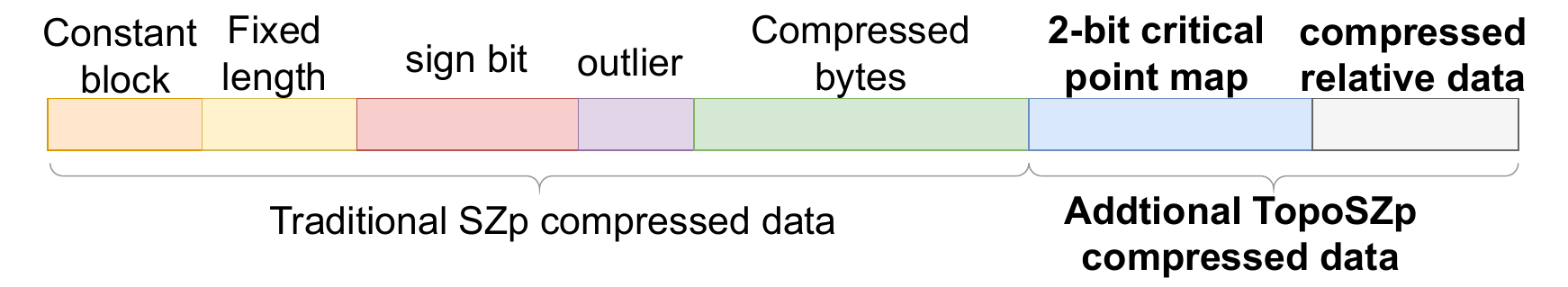}
        \vspace{-1.5em}
        \caption{Representation of compressed data in TopoSZp.}
        \label{fig:compressed_data}
    \end{figure}

    After the RP stage, the remaining compression steps (\textbf{QZ}, \textbf{B} + \textbf{LZ}, and \textbf{BE}) are applied exactly as in SZp. The resulting compressed bitstream is organized in the following order: (1) constant-block information, (2) fixed-length block metadata, (3) sign bits for all elements, (4) the first-element (outlier) value of each block, (5) the compressed byte stream generated by \textbf{QZ}, \textbf{B} + \textbf{LZ}, and \textbf{BE}, (6) 2-bit labels indicating critical vs. non-critical points, and (7) the additional bytes used to store relative rankings among critical points. The final layout of the compressed format is illustrated in Figure~\ref{fig:compressed_data}. The 2-bit critical-point labels and the relative-ordering metadata constitute the only additions beyond the standard SZp output. To further limit storage overhead, we apply the \textbf{B} + \textbf{LZ} and \textbf{BE} stages a second time—exclusively to the ordering metadata—allowing these additional bytes to be compressed efficiently. We omit \textbf{QZ} for this metadata since it is already integer-based and must remain lossless to preserve ordering relationships accurately.

    \subsection{Decompression Pipeline}
TopoSZp decompression consists of six stages: 
Fixed length byte Decoding ($\hat{\mathbf{BE}}$), inverse Decorrelation and inverse Blocking ($\hat{\mathbf{LZ}}+\hat{\mathbf{B}}$), inverse Quantization ($\hat{\mathbf{QZ}}$), additional metadata extraction ($\hat{\mathbf{MD}}$), extrema and relative order preservation ($\hat{\mathbf{CP}} + \hat{\mathbf{RP}}$), and RBF refinement for saddle points ($\hat{\mathbf{RS}}$). 
The first three stages—($\hat{\mathbf{BE}}$, $\hat{\mathbf{LZ}}+\hat{\mathbf{B}}$, and $\hat{\mathbf{QZ}}$)—follow the standard SZp decompression process, reconstructing an initial approximation $\hat{\mathcal{D}}$ that satisfies the user-defined error bound $|\mathcal{D} - \hat{\mathcal{D}}| \leq \varepsilon$. The remaining stages are unique to TopoSZp and restore the critical topological structure that may have been altered during lossy quantization. We outline these additional components in detail below.

\textbf{(1) Additional Metadata Extraction ($\hat{\mathbf{MD}}$).}
In this step, we extract the additional information stored in compressed data. This includes two components: (1) the 2-bit critical point map, and (2) the relative-order metadata compressed using an additional \textbf{B} + \textbf{LZ} and \textbf{BE} pass.
The \emph{critical-point map} is decoded directly, where each 2-bit value is mapped back to its integer class label: $\mathbf{r} = 00 \rightarrow 0$, $\mathbf{m} = 01 \rightarrow 1$, $\mathbf{s} = 10 \rightarrow 2$, $\mathbf{M} = 11 \rightarrow 3$. This restores the original per-cell classification of regular, minima, saddle, and maxima points. The \emph{relative-order metadata} is stored without quantization (to avoid further information loss) but is compressed using the standard SZp blocking and decorrelation pipeline. To recover it, we apply Fixed-Length Byte Decoding to the metadata segment ($\hat{\mathbf{BE}}_{\hat{\mathbf{MD}}}$), followed by inverse Decorrelation and inverse Blocking ($\hat{\mathbf{LZ}}_{\hat{\mathbf{MD}}}+\hat{\mathbf{B}}_{\hat{\mathbf{MD}}}$). This produces the complete lossless metadata necessary for restoring topological structure during the next correction phases. The extracted metadata is then passed to the $\hat{\mathbf{CP}} + \hat{\mathbf{RP}}$ stage, where it is used to reconstruct missing critical points and to re-establish the original ordering relationships.

\textbf{(2) Extrema and Relative Order preservation ($\hat{\mathbf{CP}} + \hat{\mathbf{RP}}$).}
\label{sec:max-min-stencil}
Using the decoded metadata, we begin by restoring any critical extrema—minima ($\mathbf{m}$) and maxima ($\mathbf{M}$)—that were lost during quantization (\textbf{Q}). For each grid point, we determine whether it was originally labeled as a critical point but appears as regular in the decompressed field; such cases represent \textit{false negatives}. If a point is identified as a lost extrema, we reconstruct it using an extrema stencil based on its saved class:
\begin{itemize}
    \item \textit{Minima stencil:} We locate all neighbors whose values are greater than or equal to the current point. The point is then updated to ensure it is strictly smaller than its neighbors by setting $\hat{\mathcal{D}}(p) = \min\limits_{q \in N(p)} \hat{\mathcal{D}}(q) - \delta \cdot \eta$, where $\eta$ is machine epsilon and $\delta$ is the rank value stored in relative-order metadata corresponding to the point $p$.
    \item \textit{Maxima stencil:} Similarly, we identify neighbors whose values are less than or equal to the target point and update them to 
    $\hat{\mathcal{D}}(p) = \max\limits_{q \in N(p)} \hat{\mathcal{D}}(q) + \delta \cdot \eta$

\end{itemize}
This procedure restores lost extrema without violating the global error bound $\varepsilon$, while the rank term $\delta$ preserves the relative ordering among extrema within the same quantization bin. Hence, $\hat{\mathbf{CP}} + \hat{\mathbf{RP}}$ ensures that (1) missing extrema are reinstated and (2) their scalar relationships remain consistent with the original field. Note that we do not introduce a saddle-based stencil. Unlike maxima and minima, saddle points are defined by more complex relationships among neighboring samples rather than by simple local extrema conditions. Designing a saddle stencil analogous to those used for maxima and minima would risk violating these neighborhood relationships, potentially introducing false positives (FP) and false types (FT). To avoid such artifacts and to preserve topological correctness, TopoSZp deliberately refrains from employing a saddle-based stencil. We therefore refine the saddle points using an RBF-based approach ($\hat{\mathbf{RS}}$) to recover and maintain saddle configurations.

\textbf{(3) RBF refinement of Saddle points 
($\hat{\mathbf{RS}})$.}
\label{sec:RBF}
Saddle points are the most fragile critical structures under lossy compression, as their classification relies on the existence of both ascending and descending directions in the local neighborhood. Even slight perturbations may break this pattern, converting a true saddle into a regular point. While the $\hat{\mathbf{CP}} + \hat{\mathbf{RP}}$ step restores extrema and preserves the relative ordering among them, preserving saddle configurations requires an additional smoothing step. To address this, we apply a lightweight radial basis function (RBF \cite{RBF}) refinement in a small neighborhood around each point (saddle point misclassified as a regular point).

Let $p$ be a saddle location and its RBF neighborhood is defined as $N_r(p) = \{\, q_i \in \Omega \mid \|q_i - p\| \le r \,\}$,
where the radius $r = \lfloor k_{\mathrm{size}}/2 \rfloor$ is determined by a small kernel size $k_{\mathrm{size}} \in \{3,5,7\}$, selected based on data  smoothness, neighbor variation, and grid resolution. We construct the interpolant
\begin{equation}
    \tilde{\mathcal{D}}(x,y) = \sum_{q_i\in N_r(p)} w_i\,\phi(\|(x,y)-(x_i,y_i)\|),
    \label{eq:RBF-interp-final}
\end{equation}
where $\phi$ is a radial basis Gaussian kernel, and weights $w_i$ are obtained from interpolation constraints $\tilde{\mathcal{D}}(q_i) = \hat{\mathcal{D}}(q_i),
     q_i\in N_r(p)$.
Evaluating~\eqref{eq:RBF-interp-final} at $p$ produces a refined    reconstruction
\begin{equation}
    \hat{\mathcal{D}}_{\mathrm{topo}}(p)
    = \tilde{\mathcal{D}}(p)
    = \sum_{q_i\in N_r(p)} \alpha_i\,\hat{\mathcal{D}}(q_i),
    \label{eq:RBF-eval}
\end{equation}
where $\alpha_i\ge 0$ and $\sum_i\alpha_i=1$. The update is therefore a convex combination of neighbor values, smoothing quantization irregularities while restoring the required saddle 
pattern:
\[
    \exists\,q_i,q_j\in N_r(p):\hat{\mathcal{D}}_{\mathrm{topo}}(q_i)>\hat{\mathcal{D}}_{\mathrm{topo}}(p),
\qquad 
\]
\[
    \exists\,q_u,q_v\in N_r(p):\hat{\mathcal{D}}_{\mathrm{topo}}(q_u)<\hat{\mathcal{D}}_{\mathrm{topo}}(p),
    \text{or vice versa.}
\]

\begin{itemize}
    \item \textit{Adaptive parameters.} To avoid manual tuning, TopoSZp estimates the RBF parameters directly from the input. The kernel width $\sigma$ is chosen in $[0.5,1.0]$ and scaled with normalized neighbor variation, larger for smooth datasets and smaller for sharp gradients. The kernel size $k_{\mathrm{size}}\in \{3,5,7\}$ is increased when global variation is low, allowing a larger support radius for stable refinement. A small tolerance $\varepsilon_{\mathrm{RBF}}=O(0.1\varepsilon)$ is used to prevent overcorrection and is adaptively tightened if local differences are smaller than the error bound. Together, these choices allow RBF refinement to be both lightweight and data-aware without requiring user input.

    \item \textit{Error bound.}  Although RBF refinement modifies reconstructed values, the update remains bounded. If the pre-refinement field satisfies $|\mathcal{D}(p)-\hat{\mathcal{D}}(p)|\le\varepsilon$ and $\mathcal{D}$ is Lipschitz-continuous \cite{Lipschitz} with constant $L$, then $|\mathcal{D}(p)-\hat{\mathcal{D}}_{\mathrm{topo}}(p)| = \varepsilon_{topo} \le\varepsilon+Lh$, where $h$ is the grid spacing.
\end{itemize}

In summary, $\hat{\mathbf{RS}}$ restores saddle structures lost during quantization, ensures local mixed-gradient behavior, and preserves topological connectivity with minimal overhead, completing the topology-aware decompression pipeline. We note, however, that $\hat{\mathbf{RS}}$ cannot recover all false-negative (FN) saddles: if quantization collapses the local gradient pattern such that all neighbors fall into the same quantization bin or the original ascent/descent relationship is entirely lost, no convex and error-bounded RBF update can recreate the sign change without exceeding the user defined error bound, and hence we deliberately avoid such situations. 


\begin{table*}[t]
    \centering
    \caption{HPC datasets and scalability of TopoSZp in terms of compression time across 1–18 threads with $\varepsilon$=1E-3. Compression time decreases with increasing thread count. The average error bound $\varepsilon_{\text{topo}}$ obtained using TopoSZp is also reported.}
    \label{tab:datasets_scale_grouped}
    \vspace{-0.5em}
    \begin{tabular}{l | c c c | c c c c c c | c}
        \toprule
        \multirow{2}{*}{\textbf{Dataset}} 
        & \multicolumn{3}{c|}{\textbf{Dataset metadata}} 
        & \multicolumn{6}{c|}{\textbf{Compression time (sec.) of TopoSZp using \# of threads (t)}} 
        & \multirow{2}{*}{%
            \begin{tabular}{@{}c@{}}
                \textbf{Err.\ Bound $\mathbf{\varepsilon}_{topo}$} \\
                (when $ \varepsilon=0.001$)
            \end{tabular}} \\
        & \textbf{\#Fields} & \textbf{Dimensions} & \textbf{Size (MB)}
        & \textbf{t = 1} & \textbf{t = 2} & \textbf{t = 4} & \textbf{t = 8} & \textbf{t = 16} & \textbf{t = 18}
        & \\
        \midrule
        \textbf{ATM}
        & 60  & $1800\times3600$ & 1483 ($\approx 1.45$GB)
        & 0.155 & 0.091 & 0.056 & 0.038 & 0.0278 & \textbf{0.0272}
        & 0.00178 \\
        \textbf{CLIMATE}
        & 90  & $768\times1152$ & 4100.5 ($\approx 4$GB)
        & 0.0289 & 0.0187 & 0.0127 & 0.0092 & \textbf{0.0078} & \textbf{0.0078}
        & 0.00180 \\
        \textbf{ICE}
        & 130 & $384\times320$ & 68.5
        & 0.0028 & 0.0017 & 0.0015 & 0.00167 & 0.00164 & \textbf{0.0016}
        & 0.00114 \\
        \textbf{LAND}
        & 176 & $192\times288$ & 40.4
        & 0.0016 & 0.00159 & 0.00157 & 0.00124 & \textbf{0.00120} & \textbf{0.00120}
        & 0.00124 \\
        \textbf{OCEAN}
        & 54  & $384\times320$ & 1096 ($\approx 1.07$GB) 
        & 0.0045 & 0.0033 & 0.00282 & 0.00281 & 0.00280 & \textbf{0.00270}
        & 0.00120 \\
        \bottomrule
    \end{tabular}
\end{table*}
        

Finally, to prevent introducing false positives (FP) or false types (FT), we track whether the refinement would generate a new or different type of critical point not present in the original critical map; if so, we suppress the correction of such saddle points to maintain topological consistency.

%% file: tex/5.performance_evaluation.tex
\section{Performance evaluation and analysis}
\label{sec:evaluate}
In this section, we evaluate the performance of TopoSZp using five scientific real-world HPC datasets.

\subsection{Experimental Settings}
We describe the environment, datasets, and different evaluation metrics used in our experiments as follows:
\begin{enumerate}
\item \textit{Environment.}
All experiments are conducted in a multi-threaded OpenMP environment on two HPC systems. On the \textbf{Argonne Bebop supercomputer}, each compute node consists of two NUMA domains, corresponding to two Intel Xeon E5-2695 v4 sockets (36 cores total) with 128 GB DRAM. Non–topology-aware compressors (SZ1.4, SZ3, ZFP, and Tthresh) are evaluated on Bebop using \textbf{18 OpenMP threads}, confined to a single NUMA domain to avoid cross-socket memory traffic. Topology-aware compressors (TopoSZ, TopoA-ZFP, TopoA-SZ3) are evaluated on the \textbf{Delta system at NCSA}.

\item \textit{Datasets.}
We evaluate TopoSZp using five floating-point scientific datasets generated by Community Earth System Model (CESM) climate simulations~\cite{cesm}, summarized in Table~\ref{tab:datasets_scale_grouped}. These datasets cover five simulation domains: atmosphere (ATM), climate (CLIMATE), ice (ICE), land (LAND), and ocean (OCEAN). CESM data are widely used benchmarks due to their large scale and complex structures.

    \item \textit{Evaluation Metrics.}
    To evaluate the performance of TopoSZp, we consider multiple metrics, including compression and decompression time, the average number of false cases (FN, FP, and FT), bit rate\footnote{Bit rate is defined as the average number of bits used to represent each data point after lossy compression. A lower bit rate corresponds to a higher compression ratio. For example, for a single-precision floating-point dataset (32 bits per value), a compression ratio of 4 implies a bit rate of 32/4=8 bits per data point.} versus topological correctness, and critical point reconstruction quality. The details of each metric are described alongside the corresponding experimental evaluations to provide a comprehensive assessment of TopoSZp’s effectiveness.

\end{enumerate}

\subsection{Performance Evaluation}
We evaluate the performance of TopoSZp across multiple metrics and datasets, and compare it against several state-of-the-art compressors.

We analyze the scalability of TopoSZp in a multi-threaded OpenMP environment (Table~\ref{tab:datasets_scale_grouped}) by varying the number of threads from 1 to 18. TopoSZp achieves \textbf{speedups of 14.2×–16.8× at 18 threads} across all evaluated datasets, corresponding to \textbf{79\%–93\% parallel efficiency} within a single NUMA domain. This demonstrates that the computational kernels of TopoSZp scale effectively with thread-level parallelism. Beyond 18 threads, additional speedups are limited due to increased synchronization overhead and cross-NUMA memory accesses \cite{NUMA-effect}. We also observe that the overall error bound using TopoSZp is bounded by $\varepsilon_{topo} \leq 2\times\varepsilon$ (see Table~\ref{tab:datasets_scale_grouped}).

    

\textbf{(1) Time Cost.} 
    We evaluate the compression and decompression time on the ATM data using the fields AEROD, CLDHGH, CLDLOW, FLDSC, and CLDMED. This metric measures the compression and decompression time of a compressor. For this analysis, we focus on topology-aware compressors, including TopoSZ, TopoA, and TopoSZp. We do not include non–topology-aware lossy compressors in this comparison, as they are optimized for compression speed and ratio and do not incorporate the additional topology-preserving steps present in TopoSZp, making direct runtime comparisons less meaningful. The goal of this experiment is to assess the efficiency of TopoSZp relative to existing topology-aware methods. Specifically, TopoSZp is designed to achieve strong topological preservation while significantly reducing compression and decompression time compared to prior topology-aware compressors. Note that, due to the high computational cost of running all topology-aware compressors over the full data suite, we perform the runtime evaluation on a representative subset consisting of five ATM fields. This subset enables a fair comparison among TopoSZ, the TopoA wrapper on the ZFP compressor (TopoA-ZFP), the TopoA wrapper on the SZ3 compressor (TopoA-SZ3), and TopoSZp within our CPU time constraints.

    \begin{figure}[!ht]
        \centering
         \vspace{-0.8em}
        \includegraphics[width=0.9\linewidth]{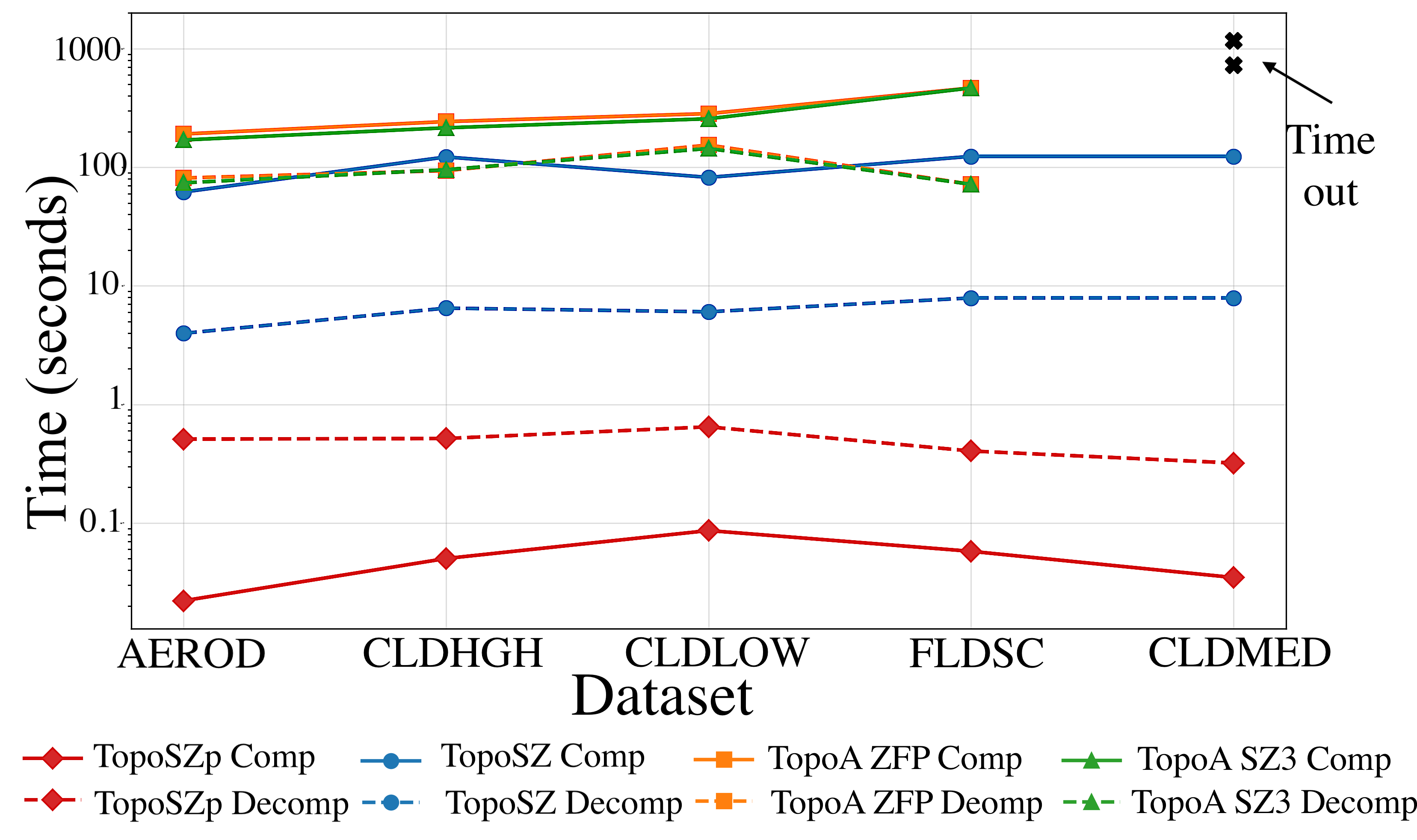}
        \vspace{-0.3em}
        \caption{Compression and Decompression time (in seconds, y-axis) taken by Topology-aware compressors (TopoSZ, TopoA (using ZFP), TopoA (using SZ3) vs TopoSZp compressor. }
        \label{fig:comp-decomp-time}
        \vspace{-0.3em}
    \end{figure}

    Fig.~\ref{fig:comp-decomp-time} presents the compression and decompression times of TopoSZp, TopoSZ, TopoA-ZFP, and TopoA-SZ3. We observe that both the compression and decompression times of TopoSZp remain consistently below one second across all data fields. Specifically, TopoSZp achieves a \textbf{1000×–-5000× speedup} in compression and a \textbf{10×–-25× speedup} in decompression compared to TopoSZ. When compared to TopoA, TopoSZp attains a \textbf{2000×–-10000× compression speedup} and a \textbf{100×–-500× decompression speedup}. These results demonstrate a substantial performance improvement over TopoSZ, TopoA-ZFP, and TopoA-SZ3. This performance advantage stems from fundamental differences in the underlying topology processing. Although TopoSZ is built on SZ1.2 and TopoA leverages ZFP and SZ3 in our experiments, their compression and decompression pipelines are dominated by the cost of topology-based descriptors, such as contour trees~\cite{carr2003topological} and persistence diagrams~\cite{edelsbrunner2002topological}. In contrast, TopoSZp relies on lightweight 2D critical point computations implemented using OpenMP during both compression and decompression. Furthermore, TopoSZp is built on the SZp compressor, which provides high compression and decompression throughput. The additional components of TopoSZp, including maxima and minima stencils and RBF-based saddle refinement, are also implemented using OpenMP-style parallelism, further contributing to its significant performance advantage.

\begin{figure}[!h]
    \centering
         \vspace{-1.2em}
    \begin{subfigure}{0.24\textwidth}
        \centering
        \includegraphics[width=\linewidth]{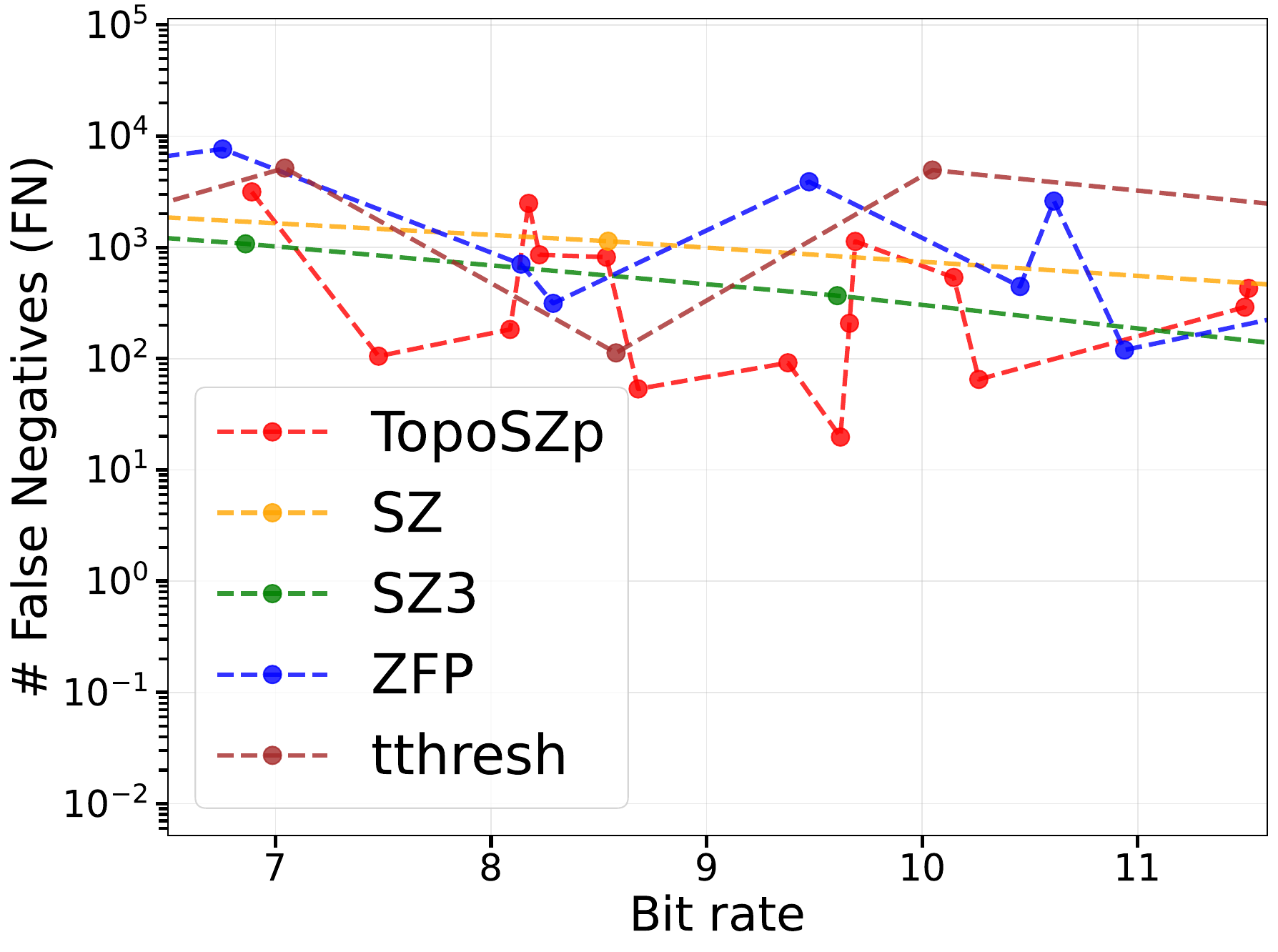}
        \vspace{-1em}
        \caption{FN}
        \label{fig:bitrate-vs-FN}
    \end{subfigure}
    \hfill
    \begin{subfigure}{0.24\textwidth}
        \centering
        \includegraphics[width=\linewidth]{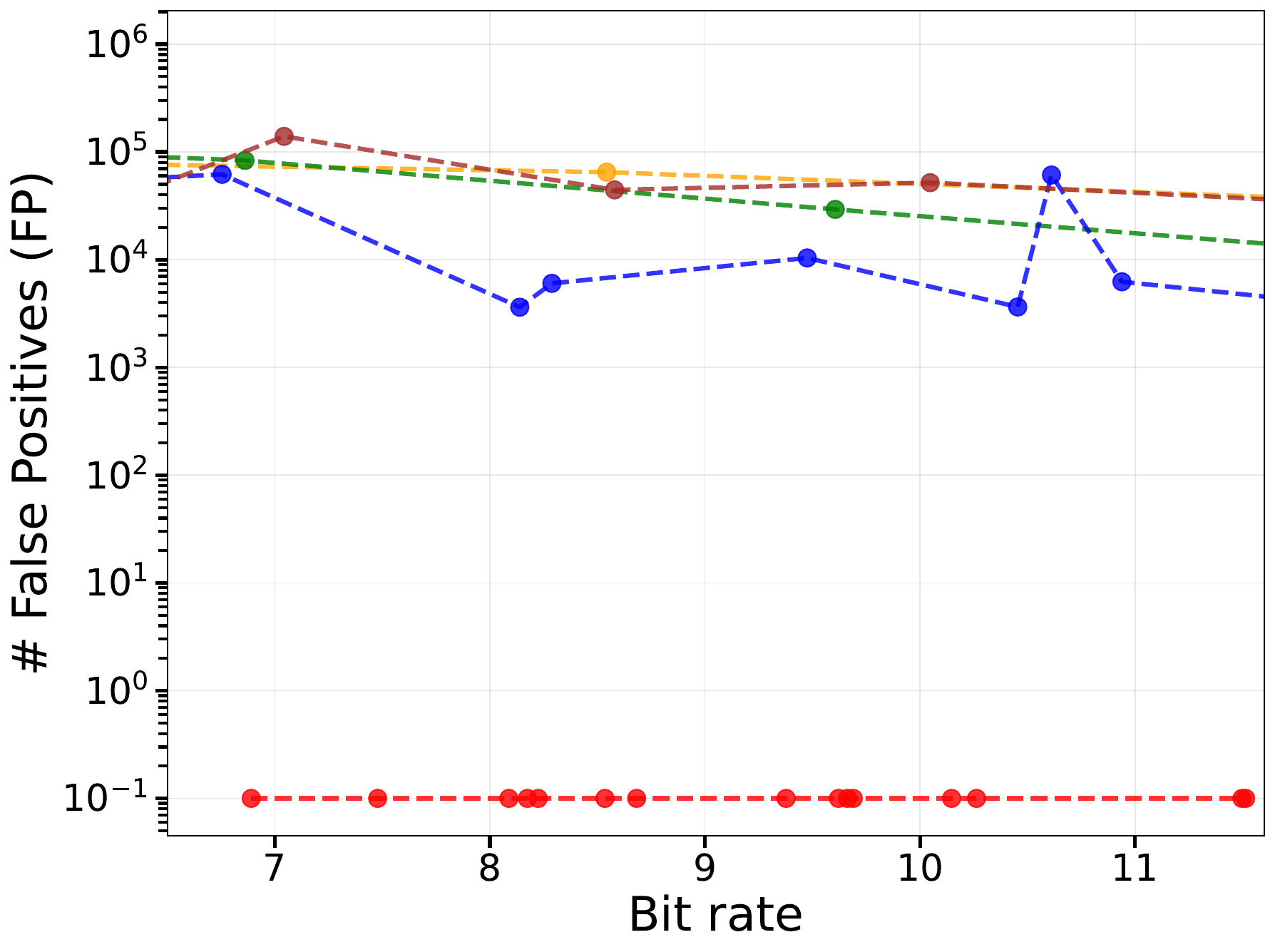}
        \vspace{-1em}
        \caption{FP}
        \label{fig:bitrate-vs-FP}
    \end{subfigure}
    \vspace{-0.2em}
    \begin{subfigure}{0.24\textwidth}
        \centering
        \includegraphics[width=\linewidth]{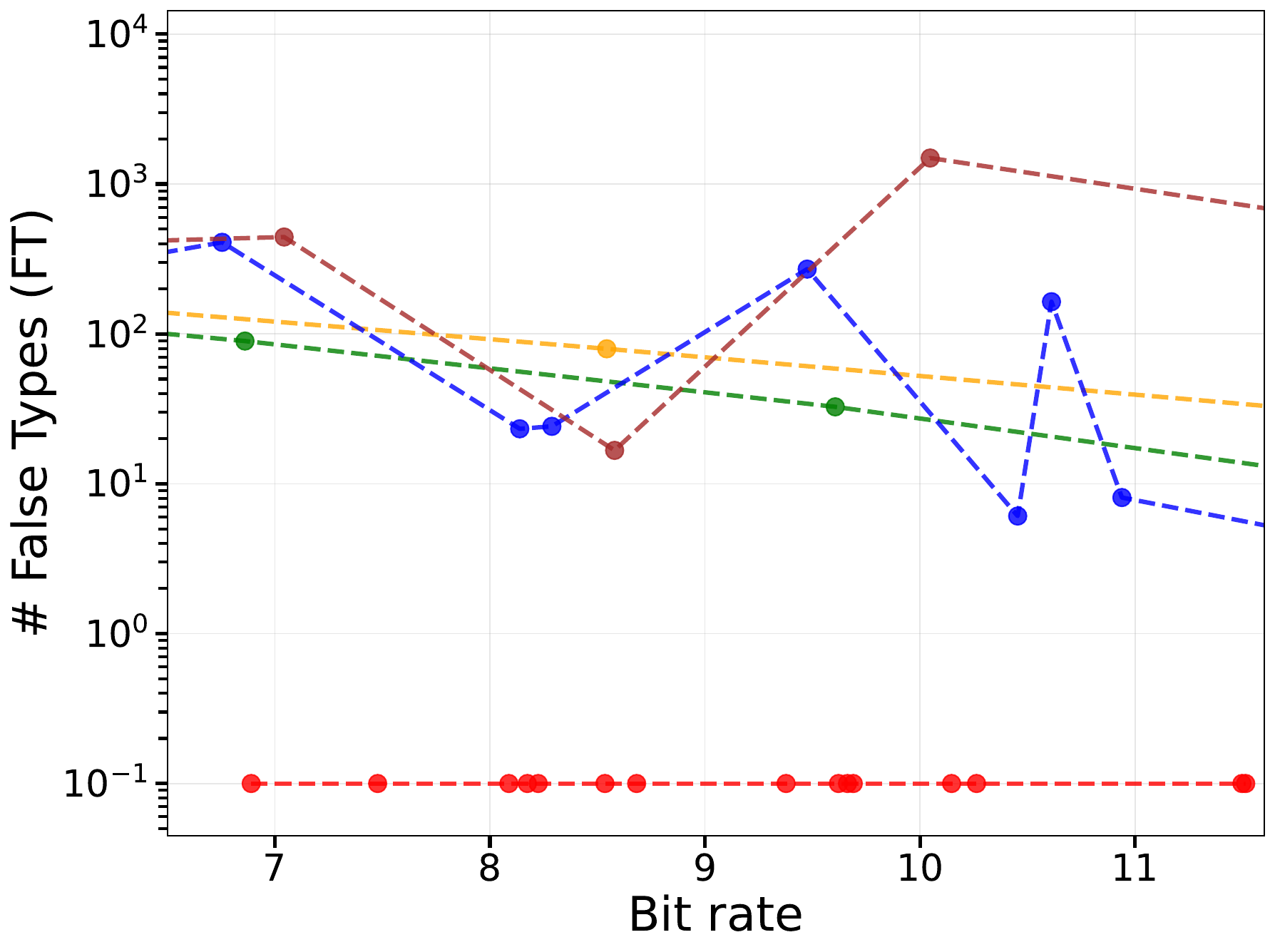}
        \vspace{-1em}
        \caption{FT}
        \label{fig:bitrate-vs-FT}
    \end{subfigure}
    \hfill
    \begin{subfigure}{0.24\textwidth}
        \centering
        \includegraphics[width=\linewidth]{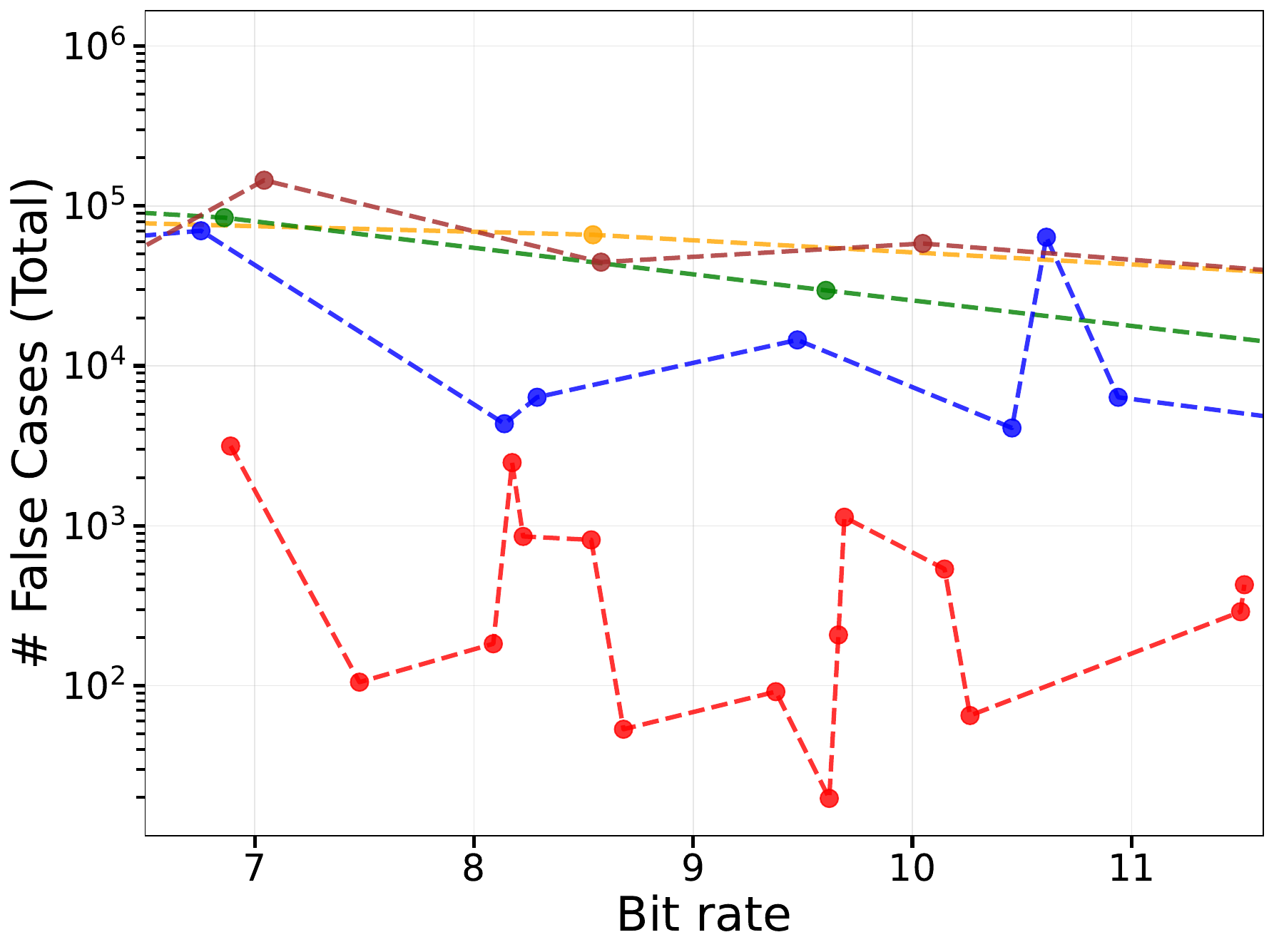}
        \vspace{-1em}
        \caption{Total}
        \label{fig:bitrate-vs-Total}
    \end{subfigure}
    \vspace{-1.8em}
    \caption{Bit rate vs average number of missed critical points for all datasets in Table~\ref{tab:datasets_scale_grouped}, i.e., FN, FP, FT, and total false cases.}
    \vspace{-1em}
    \label{fig:bitrate-vs-False}
\end{figure}

\textbf{(2) Bit Rate vs. Topological Correctness}
We present the rate–distortion plots for all error-bounded lossy compressors and TopoSZp in Fig.~\ref{fig:bitrate-vs-False}, along with their corresponding false cases. Rate–distortion is a widely used metric in the compression community, where the rate represents the average number of bits per value in the compressed data (e.g., computed as 32 divided by the compression ratio for single-precision floating-point data).
As shown in Fig.~\ref{fig:bitrate-vs-FN}, the number of false negatives produced by TopoSZp is comparable to other compressors at similar bit rates. This behavior arises because the additional topology-preserving metadata in TopoSZp reduces the compression ratio, bringing its FN counts closer to those of other compressors at equivalent rates. However, we later show that, under the same error bound, TopoSZp achieves \textbf{3×–25× fewer false negatives} than existing compressors (Table~\ref{tab:unified-eps-avg-False-Cases}). More importantly, TopoSZp completely eliminates false positives (FP) and false type (FT) errors, as shown in Figs.~\ref{fig:bitrate-vs-FP} and \ref{fig:bitrate-vs-FT}. As a result, the total number of false cases is significantly lower than that of all other lossy compressors (Fig.~\ref{fig:bitrate-vs-Total}). These results clearly demonstrate that TopoSZp provides a superior trade-off between compression efficiency and topological correctness.

 \textbf{(3) Average Number of False Cases.} 
We evaluate the total number of false negatives (FN), false positives (FP), and false types (FT) for all fields of each dataset listed in Table~\ref{tab:datasets_scale_grouped}, using multiple error-bounded lossy compressors, including TopoSZp, SZ1.2, SZ3, ZFP, and Tthresh. These metrics quantify the topological fidelity of the reconstructed data by measuring the average number of missed critical points (FN, FP and FT) introduced by lossy compression. This evaluation is performed under three user-defined absolute error bounds: 1E-3, 1E-4, and 1E-5. For each dataset and error bound, we compute the FN, FP, and FT counts for every field and then report the average values across all fields. The aggregated results are summarized in Table~\ref{tab:unified-eps-avg-False-Cases}.

\begin{table*}[!h]

\centering

\caption{Average number of False Negative (FN), False Positive (FP), False Type (FT) results for all compressors across all HPC scientific datasets and three error bounds. TopoSZp achieves \textbf{3×–25× fewer FN} and eliminates FP and FT entirely.}
\label{tab:unified-eps-avg-False-Cases}
\footnotesize

\vspace{-0.5em}
\begin{tabular}{l l | c c c | c c c | c c c}
\toprule
& & \multicolumn{3}{c|}{$\mathbf{\varepsilon_1 = 10^{{-3}}}$} & 
      \multicolumn{3}{c|}{$\mathbf{\varepsilon_2 = 10^{{-4}}}$} & 
      \multicolumn{3}{c}{$\mathbf{\varepsilon_3 = 10^{{-5}}}$}  \\
\textbf{Dataset} & \textbf{Compressor} & FN & FP & FT & FN & FP & FT & FN & FP & FT  \\
\midrule
\multirow{5}{*}{\textbf{ATM}}
 & TopoSZp & \textbf{2485.22} & \textbf{0} & \textbf{0} & \textbf{1131.70} & \textbf{0} & \textbf{0} & \textbf{428.08} & \textbf{0} & \textbf{0}  \\

 & SZ1.2 & 15161.37 & 1163.45 & 120452.35 & 7177.28 & 581.28 & 98442.80 & 2945.47 & 232.70 & 87714.67   \\

 & SZ3 & 16419.82 & 659.67 & 93707.23 & 7175.98 & 429.42 & 142447.57 & 2796.02 & 219.30 & 143711.35   \\

 & ZFP & 7651.18 & 408.32 & 61944.57 & 2608.10 & 163.92 & 61015.32 & 1104.53 & 74.25 & 45691.33
   \\

 & Tthresh & 52983.75 & 1131.55 & 151709.77 & 49062.67 & 8057.18 & 834526.70 & 45101.45 & 7919.88 & 916993.30   \\
\midrule
\multirow{5}{*}{\textbf{CLIMATE}}
 & TopoSZp & \textbf{3153.77} & \textbf{0} & \textbf{0} & \textbf{857.85} & \textbf{0} & \textbf{0} & \textbf{207.79} & \textbf{0} & \textbf{0}   \\

 & SZ1.2 & 19357.99 & 2231.22 & 85881.33 & 6568.22 & 605.57 & 43647.27 & 1725.83 & 121.97 & 26301.81   \\

 & SZ3 & 20700.43 & 3345.86 & 143989.14 & 6723.31 & 657.16 & 78041.72 & 1685.33 & 133.06 & 58877.73   \\

 & ZFP & 7354.27 & 461.60 & 32368.46 & 1466.64 & 79.76 & 17292.30 & 324.52 & 15.37 & 14945.52   \\

 & Tthresh & 42896.11 & 3327.17 & 75752.47 & 39311.03 & 7354.72 & 161627.39 & 33952.30 & 7227.48 & 185276.83   \\
\midrule
\multirow{4}{*}{\textbf{ICE}}
 & TopoSZp & \textbf{105.23} & \textbf{0} & \textbf{0} & \textbf{53.43} & \textbf{0} & \textbf{0} & \textbf{19.73} & \textbf{0} & \textbf{0}   \\

 & SZ1.2 & 471.75 & 50.62 & 22076.05 & 292.47 & 34.38 & 20026.02 & 124.48 & 12.92 & 25715.55   \\

 & SZ3 & 489.66 & 45.17 & 13571.59 & 307.42 & 32.76 & 15112.85 & 123.16 & 13.25 & 24027.65   \\
& ZFP & 313.98 & 24.16 & 6027.38 & 119.52 & 8.09 & 6235.03 & 54.80 & 2.70 & 6228.76   \\
 & Tthresh & 553.64 & 119.76 & 42833.25 & 496.96 & 129.76 & 48597.53 & 434.05 & 110.44 & 48444.03   \\
\midrule
\multirow{5}{*}{\textbf{LAND}}
 & TopoSZp & \textbf{182.93} & \textbf{0} & \textbf{0} & \textbf{91.76} & \textbf{0} & \textbf{0} & \textbf{65.09} & \textbf{0} & \textbf{0}  \\

 & SZ1.2 & 1130.70 & 138.19 & 8150.63 & 659.97 & 61.42 & 8202.40 & 404.48 & 57.46 & 7756.55   \\

 & SZ3 & 1177.47 & 155.96 & 7501.85 & 689.29 & 48.33 & 6329.20 & 404.48 & 57.81 & 5848.48  \\

 & ZFP & 707.37 & 23.26 & 3618.55 & 444.96 & 6.09 & 3639.38 & 387.51 & 3.62 & 3430.34   \\
 
 & Tthresh & 1503.43 & 376.27 & 20853.81 & 1412.53 & 402.63 & 23082.84 & 1033.22 & 260.35 & 23327.46   \\
\midrule
\multirow{5}{*}{\textbf{OCEAN}}
 & TopoSZp & \textbf{816.54} & \textbf{0} & \textbf{0} & \textbf{536.37} & \textbf{0} & \textbf{0} & \textbf{290.19} & \textbf{0} & \textbf{0}   \\

 & SZ1.2 & 4944.93 & 614.94 & 24164.33 & 3622.56 & 423.50 & 18792.06 & 2033.09 & 312.70 & 22123.41   \\

 & SZ3 & 5224.41 & 543.17 & 13720.39 & 3845.61 & 381.41 & 11107.59 & 2037.69 & 320.07 & 18481.09   \\

 & ZFP & 3889.17 & 270.59 & 10372.76 & 2132.20 & 132.26 & 9460.67 & 1190.41 & 48.48 & 8833.98   \\

 & Tthresh & 5390.02 & 1481.02 & 47942.24 & 4950.04 & 1491.11 & 51825.35 & 4474.22 & 1283.80 & 49050.67   \\

\bottomrule
\end{tabular}
\vspace{-0.6em}
\end{table*}

As shown in Table~\ref{tab:unified-eps-avg-False-Cases}, TopoSZp produces \textbf{zero false positives (FP)} and \textbf{zero false types (FT)} across all datasets and error bounds. This property is a direct consequence of the constraints quantization strategy employed by TopoSZp. 
As a result, TopoSZp cannot introduce new extrema or alter the type of an existing critical point, which prevents the creation of new critical points (FP) and incorrect critical point classifications (FT). A detailed explanation of this guarantee is provided in Section~\ref{sec:problem_B}. In addition to eliminating FP and FT, TopoSZp significantly reduces the number of false negatives \textbf{(3x--100x lesser)} compared to the other compressors. This improvement is primarily attributed to two key design components of TopoSZp. First, the maxima and minima stencil design described in Section~\ref{sec:max-min-stencil} guarantees the accurate preservation of local extrema, ensuring that FN corresponding to maxima and minima are fully resolved. Second, for the remaining FN—primarily saddle points—TopoSZp applies an RBF-based reconstruction strategy (Section~\ref{sec:RBF}). By leveraging a Gaussian kernel to reconstruct local neighborhoods, this approach effectively restores most missing saddle points while respecting the relaxed but strict error bounds. As a result, TopoSZp achieves substantially higher topological fidelity than competing error-bounded lossy compressors.

\begin{figure*}[!h]
    \centering
     \vspace{-1.2em}
     \footnotesize
    \begin{subfigure}{0.33\textwidth}
        \centering
        \includegraphics[width=\linewidth]{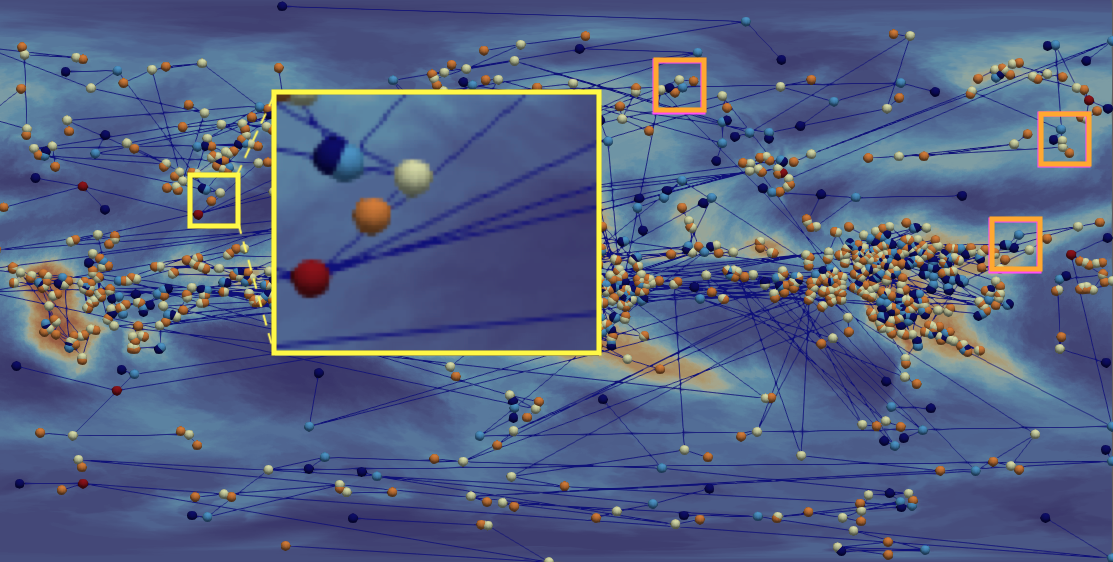}
        \caption{Original data}
        \label{fig:visualization-original}
    \end{subfigure}
    \hspace{-0.5em}
    \begin{subfigure}{0.33\textwidth}
        \centering
        \includegraphics[width=\linewidth]{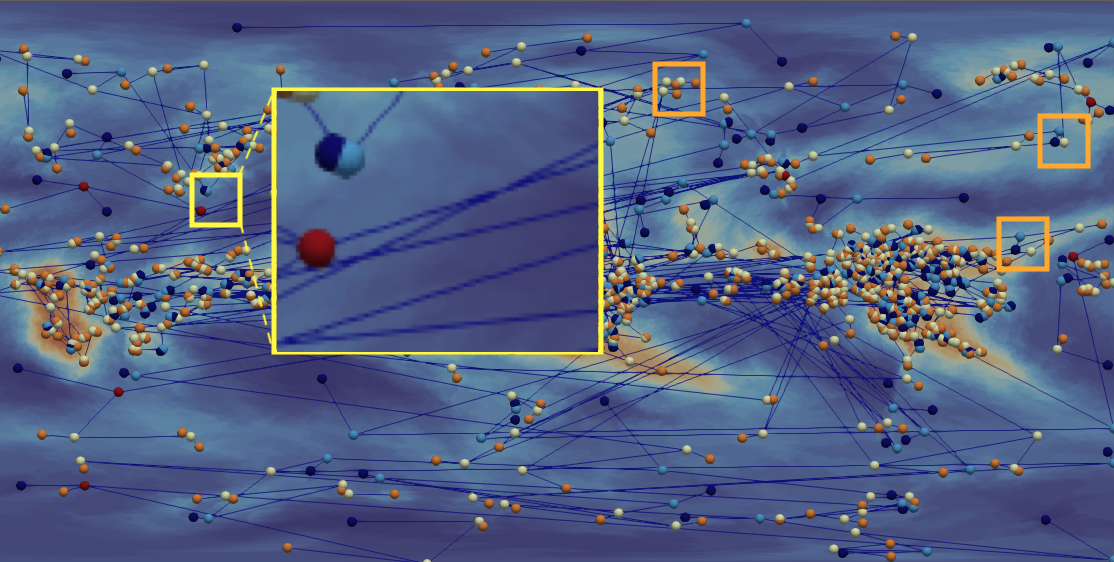}  
        \caption{SZp decompressed data}
        \label{fig:visualization-SZp}
    \end{subfigure}
    \hspace{-0.5em}
    \begin{subfigure}{0.33\textwidth}
        \centering
        \includegraphics[width=\linewidth]{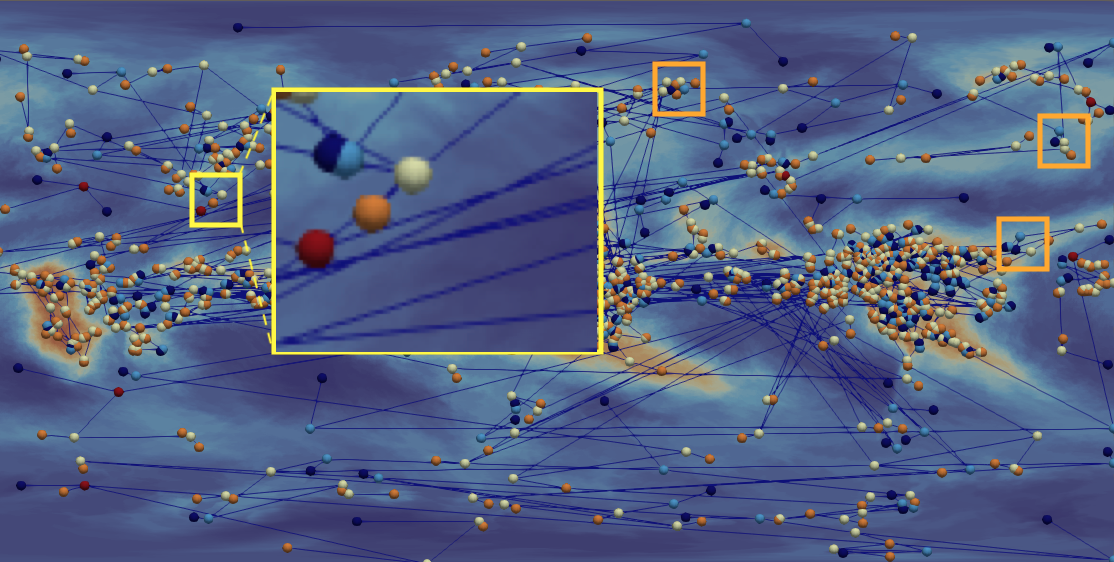} 
        \caption{TopoSZp decompressed data}
        \label{fig:visualization-TopoSZp}
    \end{subfigure}
    \vspace{-0.3em}
    \caption{Visualization of critical point reconstruction for the CLDHGH field: (a) original data, (b) SZp-decompressed data, and (c) TopoSZp-decompressed data. TopoSZp successfully preserves critical points that are missed by SZp.}
    \label{fig:visualization}
    \vspace{-1.2mm}
\end{figure*}

\textbf{(4) Critical point reconstruction Quality.}
We evaluate the data quality of the reconstructed output produced by TopoSZp through visualization of critical point reconstruction, as visualization plays a crucial role in assessing reconstruction fidelity. Specifically, we compare data decompressed using TopoSZp with data decompressed using traditional SZp (i.e., SZp decompression without applying the TopoSZp topological layers), as well as with the original data, to demonstrate TopoSZp’s ability to preserve critical points. In this experiment, we use the CLDHGH variable from the ATM data and apply compression with a user-defined absolute error bound of 1E-3. The reconstructed data are visualized to highlight differences in critical point preservation. The comparison shows (Fig.~\ref{fig:visualization}) that incorporating topology metadata in TopoSZp enables more faithful preservation of critical points than standard SZp decompression, underscoring the benefit of integrating topological information into the compression process.

 Figures~\ref{fig:visualization-original}, \ref{fig:visualization-SZp}, and \ref{fig:visualization-TopoSZp} present the original dataset, the data decompressed using SZp, and the decompressed data using TopoSZp, respectively. For each dataset, we extract critical points and their connectivity using ParaView~\cite{paraview} in conjunction with the TTK~\cite{ttk} filters. Compared to the original data, the SZp-decompressed data misses several critical points, which are highlighted by the yellow and orange squares in Figures~\ref{fig:visualization-original} and \ref{fig:visualization-SZp}. In contrast, the TopoSZp reconstruction shown in Fig.~\ref{fig:visualization-TopoSZp} \textbf{successfully preserves the critical points} that are absent in the SZp reconstruction. These results demonstrate that TopoSZp effectively captures and preserves critical points than traditional SZp compression.

%% file: tex/6.conclusion.tex
\section{Conclusion and Future Work}
\label{sec:conclude}
We propose TopoSZp, a fast, topology-aware, error-controlled lossy compressor that preserves critical points and their relationships in large-scale 2D scientific data. TopoSZp achieves this through lightweight critical point detection and optimized topological preservation strategies integrated into the compression pipeline. Extensive experiments on real-world datasets demonstrate the following key findings:
\begin{itemize}
\item TopoSZp preserves critical points and their relationships significantly better than traditional error-bounded lossy compressors, achieving \textbf{3×–100× fewer} non-preserved critical points, while strictly guaranteeing \textbf{zero false positives} and \textbf{zero false types}.
\item TopoSZp supports a \textbf{relaxed but strict error bound with theoretical guarantees}, enabling a favorable balance between compression efficiency and topological fidelity.
\item By leveraging lightweight critical point detection, extrema stencils, and RBF-based saddle refinement with multithreaded OpenMP parallelism, TopoSZp substantially reduces computational overhead, achieving \textbf{$10^2$×–$10^4$× speedup in compression} and \textbf{10×–500× speedup in decompression} compared to existing topology-aware compressors.
\end{itemize}

In future work, we plan to extend TopoSZp to three-dimensional datasets, further improve compression ratios, and generalize the framework into a modular wrapper capable of supporting multiple high-performance error-bounded lossy compressors. We also intend to extend the framework to additional computing platforms, such as GPUs.

%% file: references.bib
@article{cesm,
  title   = {The {Community Earth System Model (CESM)} large ensemble
		  project: A community resource for studying climate change
		  in the presence of internal climate variability},
  author  = {Kay, J. E. and et\ al.},
  journal = {Bulletin of the American Meteorological Society},
  volume  = {96},
  number  = {8},
  pages   = {1333--1349},
  year    = {2015}
}

@article{edelsbrunner2002topological,
  title={Topological persistence and simplification},
  author={Edelsbrunner, Herbert and Letscher, David and Zomorodian, Afra},
  journal={Discrete \& Computational Geometry},
  volume={28},
  pages={511--533},
  year={2002}
}

@article{gyulassy2008morse,
  title={Computing Morse–Smale complexes with accurate geometry},
  author={Gyulassy, Attila and others},
  journal={IEEE Transactions on Visualization and Computer Graphics},
  volume={14},
  number={6},
  pages={1614--1621},
  year={2008}
}

@article{carr2003topological,
  title={Topological computation of contour trees},
  author={Carr, Hamish and Snoeyink, Jack and Axen, Ulrike},
  journal={Computational Geometry},
  volume={24},
  pages={71--89},
  year={2003}
}

@book{milnor1963morse,
  title={Morse Theory},
  author={Milnor, John},
  publisher={Princeton University Press},
  year={1963}
}

@article{sousbie2011disPerSE,
  title={The persistent cosmic web and its filamentary structure},
  author={Sousbie, Thierry},
  journal={Monthly Notices of the Royal Astronomical Society},
  volume={414},
  year={2011}
}

@ARTICLE{helman1989visualizing,
  author={Helman, J.L. and Hesselink, L.},
  journal={IEEE Computer Graphics and Applications}, 
  title={Visualizing vector field topology in fluid flows}, 
  year={1991},
  volume={11},
  number={3},
  pages={36-46},
  doi={10.1109/38.79452}}

@article{edelsbrunner2003morse,
  title={Hierarchical Morse–Smale complexes for piecewise linear 2-manifolds},
  author={Edelsbrunner, Herbert and Harer, John and Zomorodian, Afra},
  journal={Discrete \& Computational Geometry},
  volume={30},
  pages={87--107},
  year={2003}
}

@article{Muszynski2019AR,
  author    = {Grzegorz Muszynski and Karthik Kashinath and Vitaliy Kurlin and Michael Wehner and Prabhat},
  title     = {Topological Data Analysis and Machine Learning for Recognizing Atmospheric River Patterns in Large Climate Datasets},
  journal   = {Geoscientific Model Development},
  volume    = {12},
  pages     = {613--628},
  year      = {2019},
  doi       = {10.5194/gmd-12-613-2019}
}

@article{Carr2003ContourTrees,
  author  = {Carr, Hamish and Snoeyink, Jack and Axen, Ulrike},
  title   = {Computing Contour Trees in All Dimensions},
  journal = {Computational Geometry},
  volume  = {24},
  number  = {2},
  pages   = {75--94},
  year    = {2003},
  doi     = {10.1016/S0925-7721(02)00094-7}
}

@inproceedings{feature-tracking,
author = {Corcoran, Padraig and Jones, Christopher B.},
title = {Robust tracking of objects with dynamic topology},
year = {2018},
isbn = {9781450358897},
publisher = {Association for Computing Machinery},
address = {New York, NY, USA},
doi = {10.1145/3274895.3274922},
booktitle = {Proceedings of the 26th ACM SIGSPATIAL International Conference on Advances in Geographic Information Systems},
pages = {428–431},
numpages = {4},
location = {Seattle, Washington},
series = {SIGSPATIAL '18}
}

@article{TopoSZ,
author = {Yan, Lin and Liang, Xin and Guo, Hanqi and Wang, Bei},
title = {TopoSZ: Preserving Topology in Error-Bounded Lossy Compression},
year = {2024},
issue_date = {Jan. 2024},
publisher = {IEEE Educational Activities Department},
address = {USA},
volume = {30},
number = {1},
issn = {1077-2626},
doi = {10.1109/TVCG.2023.3326920},
journal = {IEEE Transactions on Visualization and Computer Graphics},
month = jan,
pages = {1302–1312},
numpages = {11}
}

@article{Soler2018TopologicallyCL,
  title={Topologically Controlled Lossy Compression},
  author={Maxime Soler and M{\'e}lanie Plainchault and Bruno Conche and Julien Tierny},
  journal={2018 IEEE Pacific Visualization Symposium (PacificVis)},
  year={2018},
  pages={46-55},
}

@ARTICLE{TopoA,
author={Gorski, Nathaniel and Liang, Xin and Guo, Hanqi and Yan, Lin and Wang, Bei},
journal={ IEEE Transactions on Visualization \& Computer Graphics },
title={{ A General Framework for Augmenting Lossy Compressors With Topological Guarantees }},
year={2025},
volume={31},
number={06},
ISSN={1941-0506},
pages={3693-3705},
doi={10.1109/TVCG.2025.3567054},
publisher={IEEE Computer Society},
address={Los Alamitos, CA, USA},
month=jun}

@article{Fugacci,
author = {Fugacci, Ulderico and Kerber, Michael and Rolle, Alexander},
title = {Compression for 2-parameter persistent homology},
year = {2023},
issue_date = {Feb 2023},
publisher = {Elsevier Science Publishers B. V.},
address = {NLD},
volume = {109},
number = {C},
issn = {0925-7721},
doi = {10.1016/j.comgeo.2022.101940},
journal = {Comput. Geom. Theory Appl.},
month = feb,
numpages = {28}
}

@misc{TFZ,
      title={TFZ: Topology-Preserving Compression of 2D Symmetric and Asymmetric Second-Order Tensor Fields}, 
      author={Nathaniel Gorski and Xin Liang and Hanqi Guo and Bei Wang},
      year={2025},
      eprint={2508.09235},
      archivePrefix={arXiv},
      primaryClass={cs.GR},
      url={https://arxiv.org/abs/2508.09235}, 
}

@INPROCEEDINGS{Sign-of-determinant,
  author={Xia, Mingze and Di, Sheng and Cappello, Franck and Jiao, Pu and Zhao, Kai and Liu, Jinyang and Wu, Xuan and Liang, Xin and Guo, Hanqi},
  booktitle={2024 IEEE 40th International Conference on Data Engineering (ICDE)}, 
  title={Preserving Topological Feature with Sign-of-Determinant Predicates in Lossy Compression: A Case Study of Vector Field Critical Points}, 
  year={2024},
  volume={},
  number={},
  pages={4979-4992},
  doi={10.1109/ICDE60146.2024.00378}}

@article{MEYERBASE2005383,
title = {Medical image compression using topology-preserving neural networks},
journal = {Engineering Applications of Artificial Intelligence},
volume = {18},
number = {4},
pages = {383-392},
year = {2005},
issn = {0952-1976},
doi = {https://doi.org/10.1016/j.engappai.2004.10.004},
author = {Anke Meyer-Bäse and Karsten Jancke and Axel Wismüller and Simon Foo and Thomas Martinetz}
}

@INPROCEEDINGS{SZ1.4,
  author={Tao, Dingwen and Di, Sheng and Chen, Zizhong and Cappello, Franck},
  booktitle={2017 IEEE International Parallel and Distributed Processing Symposium (IPDPS)}, 
  title={Significantly Improving Lossy Compression for Scientific Data Sets Based on Multidimensional Prediction and Error-Controlled Quantization}, 
  year={2017},
  volume={},
  number={},
  pages={1129-1139},
  doi={10.1109/IPDPS.2017.115}}

@INPROCEEDINGS{SZ2,
  author={Liang, Xin and Di, Sheng and Tao, Dingwen and Li, Sihuan and Li, Shaomeng and Guo, Hanqi and Chen, Zizhong and Cappello, Franck},
  booktitle={2018 IEEE International Conference on Big Data (Big Data)}, 
  title={Error-Controlled Lossy Compression Optimized for High Compression Ratios of Scientific Datasets}, 
  year={2018},
  volume={},
  number={},
  pages={438-447},
  doi={10.1109/BigData.2018.8622520}}

@ARTICLE{SZ3,
  author={Liang, Xin and Zhao, Kai and Di, Sheng and Li, Sihuan and Underwood, Robert and Gok, Ali M. and Tian, Jiannan and Deng, Junjing and Calhoun, Jon C. and Tao, Dingwen and Chen, Zizhong and Cappello, Franck},
  journal={IEEE Transactions on Big Data}, 
  title={SZ3: A Modular Framework for Composing Prediction-Based Error-Bounded Lossy Compressors}, 
  year={2023},
  volume={9},
  number={2},
  doi={10.1109/TBDATA.2022.3201176}}

@inproceedings{cuSZp,
author = {Huang, Yafan and Di, Sheng and Yu, Xiaodong and Li, Guanpeng and Cappello, Franck},
title = {cuSZp: An Ultra-fast GPU Error-bounded Lossy Compression Framework with Optimized End-to-End Performance},
year = {2023},
isbn = {9798400701092},
publisher = {Association for Computing Machinery},
address = {New York, NY, USA},
url = {https://doi.org/10.1145/3581784.3607048},
doi = {10.1145/3581784.3607048},
booktitle = {Proceedings of the International Conference for High Performance Computing, Networking, Storage and Analysis},
articleno = {43},
numpages = {13},
location = {Denver, CO, USA},
series = {SC '23}
}

@misc{hoSZp,
      title={HoSZp: An Efficient Homomorphic Error-bounded Lossy Compressor for Scientific Data}, 
      author={Tripti Agarwal and Sheng Di and Jiajun Huang and Yafan Huang and Ganesh Gopalakrishnan and Robert Underwood and Kai Zhao and Xin Liang and Guanpeng Li and Franck Cappello},
      year={2024},
      eprint={2408.11971},
      archivePrefix={arXiv},
      primaryClass={cs.DC},
      url={https://arxiv.org/abs/2408.11971}, 
}

@INPROCEEDINGS{SZOps,
  author={Agarwal, Tripti and Di, Sheng and Huang, Jiajun and Huang, Yafan and Gopalakrishnan, Ganesh and Underwood, Robert and Zhao, Kai and Liang, Xin and Li, Guanpeng and Cappello, Franck},
  booktitle={SC24-W: Workshops of the International Conference for High Performance Computing, Networking, Storage and Analysis}, 
  title={SZOps: Scalar Operations for Error-bounded Lossy Compressor for Scientific Data}, 
  year={2024},
  volume={},
  number={},
  pages={260-269},
  doi={10.1109/SCW63240.2024.00042}}

@INPROCEEDINGS{hZCCL,
  author={Huang, Jiajun and Di, Sheng and Yu, Xiaodong and Zhai, Yujia and Liu, Jinyang and Jian, Zizhe and Liang, Xin and Zhao, Kai and Lu, Xiaoyi and Chen, Zizhong and Cappello, Franck and Guo, Yanfei and Thakur, Rajeev},
  booktitle={SC24: International Conference for High Performance Computing, Networking, Storage and Analysis}, 
  title={hZCCL: Accelerating Collective Communication with Co-Designed Homomorphic Compression}, 
  year={2024},
  volume={},
  number={},
  pages={1-15},
  doi={10.1109/SC41406.2024.00110}}

@ARTICLE{ZFP,
  author={Lindstrom, Peter},
  journal={IEEE Transactions on Visualization and Computer Graphics}, 
  title={Fixed-Rate Compressed Floating-Point Arrays}, 
  year={2014},
  volume={20},
  number={12},
  pages={2674-2683},
  doi={10.1109/TVCG.2014.2346458}}

@article{MGARD,
title = {MGARD: A multigrid framework for high-performance, error-controlled data compression and refactoring},
journal = {SoftwareX},
volume = {24},
pages = {101590},
year = {2023},
issn = {2352-7110},
doi = {https://doi.org/10.1016/j.softx.2023.101590},
author = {Qian Gong and Jieyang Chen and Ben Whitney and Xin Liang and Viktor Reshniak and Tania Banerjee and Jaemoon Lee and Anand Rangarajan and Lipeng Wan and Nicolas Vidal and Qing Liu and Ana Gainaru and Norbert Podhorszki and Richard Archibald and Sanjay Ranka and Scott Klasky}
}

@ARTICLE{TTHRESH,
  author={Ballester-Ripoll, Rafael and Lindstrom, Peter and Pajarola, Renato},
  journal={IEEE Transactions on Visualization and Computer Graphics}, 
  title={TTHRESH: Tensor Compression for Multidimensional Visual Data}, 
  year={2020},
  volume={26},
  number={9},
  doi={10.1109/TVCG.2019.2904063}}

@article{Lorenzo_prediction,
author = {Ibarria, Lawrence and Lindstrom, Peter and Rossignac, Jarek and Szymczak, Andrzej},
title = {Out-of-core compression and decompression of large n-dimensional scalar fields},
journal = {Computer Graphics Forum},
volume = {22},
number = {3},
pages = {343-348},
doi = {https://doi.org/10.1111/1467-8659.00681},
eprint = {https://onlinelibrary.wiley.com/doi/pdf/10.1111/1467-8659.00681},
year = {2003}
}

@incollection{Huffman,
  author = {Puneet Mangla},
  title = {{Implementing Huffman Encoding for Lossless Compression}},
  booktitle = {PyImageSearch},
  editor = {Puneet Chugh and Susan Huot and Piyush Thakur},
  year = {2025},
  url = {https://pyimg.co/t2swr},
}

@misc{gzip1992,
  title        = {{GZIP}: GNU zip compression utility},
  author       = {Deutsch, Peter and Gailly, Jean-loup},
  howpublished = {\url{https://www.gnu.org/software/gzip/}},
  year         = {1992},
  note         = {Accessed: [date]}
}

@incollection{Lipschitz,
  author    = {Donald Estep},
  title     = {Lipschitz Continuity},
  booktitle = {Practical Analysis in One Variable},
  publisher = {Springer},
  year      = {2002},
  pages     = {83--98},
}

@book{RBF,
  title     = {Radial Basis Functions: Theory and Implementations},
  author    = {Buhmann, Martin D.},
  year      = {2003},
  publisher = {Cambridge University Press}
}

@book{paraview,
author = {Ayachit, Utkarsh},
title = {The ParaView Guide: A Parallel Visualization Application},
year = {2015},
isbn = {1930934300},
publisher = {Kitware, Inc.},
address = {Clifton Park, NY, USA}
}

@article{ttk,
  title={The topology toolkit},
  author={Tierny, Julien and Favelier, Guillaume and Levine, Joshua A and Gueunet, Charles and Michaux, Michael},
  journal={IEEE transactions on visualization and computer graphics},
  volume={24},
  number={1},
  pages={832--842},
  year={2017},
  publisher={IEEE}
}

@inproceedings{NUMA-effect,
author = {Blagodurov, Sergey and Zhuravlev, Sergey and Dashti, Mohammad and Fedorova, Alexandra},
title = {A case for NUMA-aware contention management on multicore systems},
year = {2011},
publisher = {USENIX Association},
address = {USA},
booktitle = {Proceedings of the 2011 USENIX Conference on USENIX Annual Technical Conference},
location = {Portland, OR},
series = {USENIXATC'11}
}
